\documentclass[twocolumn,tighten,times]{aastex701}

\usepackage{amsmath,amssymb}
\usepackage[export]{adjustbox}% http://ctan.org/pkg/adjustbox
\usepackage{stackengine}
\usepackage{todonotes}
\usepackage{xcolor}

\definecolor{DarkBlue}{rgb}{0.1,0.1,0.5}
\definecolor{DarkGreen}{rgb}{0.1,0.5,0.1}

\newcommand{\lira}{_{\rm Lira}}
\newcommand{\coh}{_{\rm Coh}}
\newcommand{\chandra}{{\it Chandra}}
\newcommand{\LamMC}{\tilde\Lambda^{(t)}}
\newcommand{\LamMCone}{\tilde\Lambda^{(1)}}
\newcommand{\ZMC}{Z^{(t)}}
\newcommand{\PsiMC}{\Psi^{(t)}}
\newcommand{\thetaMC}{\theta^{(t)}}
\newcommand{\betaMC}{\beta^{(t)}}
\newcommand{\cur}{^{(s)}}
\newcommand{\prev}{^{(s-1)}}
\newcommand{\base}{_{\rm b}}
\newcommand{\Ex}{\hat{\mathbb E}}
\newcommand{\hatLamtilde}{\Ex(\tilde\Lambda \mid Y)}
\newcommand{\hatZ}{\Ex(Z \mid Y)}

\usepackage{placeins}

\shorttitle{Estimating the boundaries of complex structures}
\shortauthors{McKeough et al.}

\graphicspath{{./}{figures/}}

\begin{document}

\title{The LIRA-Ising Model: Estimating the boundaries of irregularly shaped X-ray sources} %

    \author[0000-0002-0786-7307]{Kathryn McKeough}
    \affiliation{Department of Statistics, Harvard University, Cambridge, Massachusetts 02138, USA}
    \email{katy.mckeough@gmail.com}
    \author[0000-0002-3869-7996]{Vinay L.\ Kashyap}
    \affiliation{Center for Astrophysics $|$ Harvard \& Smithsonian, Cambridge MA 02138 USA}
    \email{vkashyap@cfa.harvard.edu}
    \author[0000-0002-0905-7375]{Aneta Siemiginowska}
    \affiliation{Center for Astrophysics $|$ Harvard \& Smithsonian, Cambridge MA 02138 USA}
    \email{asiemiginowska@cfa.harvard.edu}
    \author[0000-0002-0816-331X]{David A.\ van Dyk}
    \affiliation{Department of Mathematics, Imperial College London, London SW7 2AZ UK}
    \email{d.van-dyk@imperial.ac.uk}
    \author{Shihao Yang}
    \affiliation{H.\ Milton Stewart School of Industrial and Systems Engineering, Georgia Tech, Atlanta 30332, USA}
    \email{shihao.yang@isye.gatech.edu}
    \author[0000-0003-3687-0385]{Xiao-Li Meng}
    \affiliation{Department of Statistics, Harvard University, Cambridge, Massachusetts 02138, USA}
    \email{xlmeng@g.harvard.edu}
    \author[0009-0007-2168-8261]{Brendan Martin}
    \affiliation{Department of Mathematics, Imperial College London, London SW7 2AZ UK}
    \email{brendanthomasmartin@gmail.com}
    \author[0000-0001-8952-676X]{Andreas Zezas}
    \affiliation{Department of Physics, University of Crete, GR-71003 Heraklion, Greece}
    \email{azezas@physics.uoc.gr}
    \correspondingauthor{Aneta Siemiginowska}
    \email{asiemiginowska@cfa.harvard.edu}



\begin{abstract}
Mapping the boundary of an extended source is a key step in the study of its morphology. The background contamination and statistical fluctuations of typical astronomical images make this a challenging statistical task, particularly for X-ray images with low surface brightness.  We develop a three-step Bayesian procedure to identify the boundaries of irregularly shaped sources. We first apply a Bayesian multiscale reconstruction algorithm known as LIRA to obtain posterior pixelwise probability distributions of the source intensity that properly account for known structures, astrophysical background, and the effect of the telescope point spread function.  Next, we adopt an Ising model to group pixels with similar intensities into cohesive regions corresponding to background and source. Finally, the boundary is derived on the basis of the most likely aggregation of pixels into the source region. Because the overall model combines LIRA and the Ising model, we call it LIRA-Ising. We verify the proposed method using a set of simulation studies. We then apply it to the \chandra\ X-ray Observatory images of two high redshift quasars, PKS J1421$-$0643 and 0730$+$257, to determine the extent and morphology of X-ray jets. Our method shows a uniform X-ray surface brightness of PKS\,J1421$-$0643 jet, and identifies knotty structure in the X-ray jet of 0730+257.
\end{abstract}

\keywords{\uat{Interdisciplinary astronomy}{804} --- \uat{Astrostatistics}{1882} --- \uat{Algorithms}{1883} ---\uat{Astrostatistics techniques}{1886} ---\uat{Deconvolution}{1910} --- \uat{Computational methods}{1965} --- \uat{Bayesian statistics}{1900}}


\section{Introduction}

High angular resolution images obtained by the \chandra\, X-ray Observatory (\chandra) provide an unprecedented view of the X-ray Universe \cite[e.g., see][for science review, and also images on the \chandra\ website\footnote{\url{https://chandra.harvard.edu/}}]{ChandraBook}. These X-ray images trace complex structures associated with high energy events, e.g., supernova explosions, galaxy mergers, or relativistic jets, and provide important, otherwise not accessible, information for understanding the physics of many astronomical objects. For example, the X-ray morphology of Supernova Remnants (SNR) leads to a reconstruction of the exploded progenitor star structure \cite[see, e.g.][]{DeLaney2010,Vink2012}, while the `rings' and 'bubbles' imprinted in X-ray clusters indicate past episodes of a supermassive black hole activity compressing and shocking a hot intercluster medium (ICM) \citep[e.g.][]{Fabian2006,Fabian2012}. However, the observed X-ray morphology is typically diffuse and irregularly shaped, and the observed images are subject to imperfections of the telescope and detectors, and to fluctuations due to Poisson count noise. Therefore,  X-ray images cannot easily be modeled, and defining, for example, the boundaries of interesting structures is a difficult task. 

\chandra\, X-ray images offer the highest angular resolution available to date, but are susceptible to instrumental effects that are not easy to `remove'. Specifically, the observed images are blurred by the mirrors as quantified by an energy dependent point spread function (PSF) that varies across the field of view. 
It is thus challenging to resolve detailed features, and to determine the intensities of features with low angular separation, especially those with large intensity differences, or to determine the boundary of extended faint structures without clearly visible edges.

%The goal of this project
Our goal is to develop a method to map the boundary of an extended structure that is not easily deciphered in low-count Poisson images. We present a new 
principled statistical method for tracing the boundary of an extended source in a low-count X-ray image. The method can be viewed as a post-processing of the joint posterior distribution for the pixelized source intensities and is based on a model for the boundary that favors spatial cohesion.
We provide details of a three-step procedure that estimates the boundary and also returns a 
posterior probability for each pixel being associated with the extended source enclosed by the fitted boundary. These posterior pixel probabilities provide a principled quantification of uncertainty of the extended source region. 

Mapping source boundaries can influence future analyses and scientific inference. One approach is to define boundaries by hand, which in X-ray images %can be tedious and 
can induce errors caused by human subjectivity. In some situations, high-resolution images from other wavebands, i.e., radio or optical, can be used to specify regions and measure the X-ray intensities in such pre-defined regions.
This strategy, however, can be
sensitive to the boundary shape and size \citep[e.g., as in][]{McKeough2016}. 
Another approach involves using scientific models to describe the morphology of these objects \citep[e.g.,][]{Plsek2024}. However, data sparsity means evaluating these models can be tricky, and it is often these scientific models that we aim to understand. For these reasons, we prefer a data-based automated method. 

Data-based approaches for source detection and boundary algorithms are widely used in astronomy. However, no method currently exists that can accommodate the hurdles that come with sparse, X-ray images. Wavelet-based approaches, including \texttt{wavdetect}, are useful for detecting point sources in a noisy environment, including X-ray images \citep{Starck2002,Freeman2002}. \citet{Vikhlinin1998} uses a matched filter technique to detect sources in X-ray images. BASCS \citep{Jones2014} is a Bayesian technique that uses spatial and spectral data to probabilistically separate photons into two or more overlapping point sources; eBASCS \citep{meye:etal21} is an extension of BASCS that also 
leverages information from temporal variations. %uses arrival times. 
These approaches have proven useful for detecting and reconstructing point sources, but are not designed for cases with extended emission. Other attempts to model more complicated structure have been developed in high signal to noise settings. However, these approaches do not perform well with sparse, low-signal, X-ray images. Scientists have used adaptive binning to smooth images of galaxy clusters to help map physical parameters \citep{Sanders2001, Sanders2006}. \citet{Picquenot2019} developed a segmentation method for extended sources, but it requires high photon counts as well as spectral homogeneity, an requirement not guaranteed in all observations. 

A method that works with complex extended sources is adaptive kernel smoothing, which creates a smooth representation of the input data. However, it remains unclear how this can apply for scientific purposes \citep{Ebeling2006}. Spatial field reconstruction using a Gaussian Markov random field to model extended sources has performed well in survey data of galaxies and could be extended to X-ray images \citep{Gonzalez-Gaitan2019}. The technique known as {\tt{vtpdetect}} is widely implemented, but is limited by computation cost and the need for global thresholding \citep{Ebeling1993}, and furthermore does not take the PSF into account. 

Methods similar to seeded region growing have been adapted to capture irregular shapes, but only in high-signal images \citep{Bertin1996}.  Typical machine learning techniques, such as morphological snakes \citep{Marquez-Neila2014} or seeded region growing \citep{Adams1994}, tend to have trouble segmenting sparse images. Such techniques rely on large amounts of data and do not {directly provide estimates of uncertainty}. {SRGonG \citep{fan:etal:23} is a penalized likelihood method that uses greedy optimization based on seeded region growing to directly segment event lists. Auto-BUQ extends SRGonG to quantify uncertainty on the boundary between two segments \citep{wang:etal:24}. However, neither SRGonG nor Auto-BUQ account for instrumental effects such as point spread functions or exposure maps.} All of these techniques have weaknesses in addressing boundaries of low count, diffuse sources and therefore a specialized approach is required.

Bayesian techniques are useful for dealing with X-ray images since the process of collecting photons is inherently probabilistic. Markov chain Monte Carlo (MCMC) algorithms allow us to explore the full posterior distribution of the parameters and thus obtain uncertainty measurements on our estimates of quantities of interest. Furthermore, a Bayesian reconstruction algorithm simplifies parameter inference by removing ambiguity about stopping rules, as well as providing estimates for all of the model parameters simultaneously. To elicit small-scale detail in noisy images our three-step procedure first uses Low-counts Image Reconstruction and Analysis \citep[LIRA\footnote{LIRA is an open-source package available on GitHub: \\ \textbf{R}: \url{https://github.com/astrostat/LIRA}; Python: \url{https://github.com/astrostat/pylira}},][]{Esch2004,Connors2007,Donath2022}. LIRA is a fully Bayesian model and MCMC algorithm designed to simultaneously model the structure of an observed image at multiple scales. This multiscale structure captures the residual emission in excess of a specified baseline model (e.g., composed of a uniform background and known point sources). LIRA is effective in eliciting the details of extended sources within this multiscale component. It can also be used in the presence of a PSF and is especially useful when the structure of interest is on the same scale or smaller than that of the PSF. It was previously used in the detection of source components of extragalactic jets \citep{Stein2015, McKeough2016, Reddy2021, Reddy2023}. LIRA has been used to recover the detailed structure of the supernova remnant SN\,1987\,A across multiple observations \citep{Kashyap2017}, diffuse hot halos in nearby galaxies \citep{Borlaff2024,Borlaff2024b,Chamba2025},  and also to resolve small-scale features in  complex nuclear regions of radio galaxies \citep[e.g.,][]{Fabbiano2019}.

However, while LIRA is capable of computing principled uncertainty intervals on the intensities at every pixel, it cannot identify {\sl connected regions} in the field of view.  Here, we extend the parameterization of LIRA by incorporating an indicator variable that identifies the pixels that are associated with the extended source. This involves the introduction of a ``cohesion prior distribution'' that uses an Ising model to quantify the expectation that the pixels associated with an extended source be spatially cohesive. The Ising model and its generalized version, the Potts model, were initially proposed to model spin states in ferromagnetic materials \citep{ising1925, potts_1952}. It has since been widely used in image segmentation in many fields, including medical images and sonar images of the Earth \citep{Bentrem2010, Mignotte2000a}. We expand upon these applications of the Ising model 
with our proposed Bayesian method for segmenting low signal-to-noise images. Because LIRA's multiscale strategy has proven useful in practice, we aim to maintain its multiscale prior specification. Thus, we layer a new prior distribution based on the Ising model atop the LIRA prior, thus adding the expectation of spatial cohesion to the overall model. We emphasize that this is a fully Bayesian approach in that the two priors form a particular mathematical factorization of the joint prior on the expanded parameterization.  This allows us to introduce two scientifically meaningful components to the overall model (multiscale structure and spatial cohesion) and to do so in a statistically principled and computationally practical manner.  

Because our proposed method combines the LIRA and Ising models, we refer to it as LIRA-Ising. The procedure consists of three steps: (1) fitting the standard LIRA model, (2) using the posterior sample obtained with LIRA to fit the cohesion prior distribution via an Ising model on the variable that indicates which pixels are associated with the extended source, and (3) optimizing the overall model over the indicator variable to obtain an estimate of the boundary. We apply the proposed method to examine the X-ray images of extragalactic jets associated with high redshift ($z>2 $) quasars, obtained with the \textit{Chandra} telescope and presented in \cite{McKeough2016}. Understanding the morphology of jets is essential in understanding the underlying physics that creates them.

In Section \ref{sec:ising:method}, we formalize the details of the LIRA-Ising model and show how our three-step fully Bayesian image segmentation procedure  can be used to estimate the boundary of an extended source. Mathematical details of the model appear in Appendix~\ref{app:joint-post}
and the computational methods used to fit the model are described in Appendices~\ref{app:gibbs} and \ref{app:Zmargin}. Sections \ref{sec:ising:validate} and \ref{sec:ising:results} validate the method through several simulation studies and its application to two extragalactic jets, respectively.

\section{Model \& Inference}
\label{sec:ising:method}

\subsection{LIRA: Low-count Image Reconstruction and Analysis}
\label{sec:lira}

\citet{Esch2004} proposed a fully Bayesian reconstruction technique for low-count high-energy astrophysical image analysis. Using a Gibbs sampler to simulate from the posterior distribution of the image allows this method to provide principled {pixel-wise} uncertainty quantification.  Building upon \citeauthor{Esch2004}'s method, \citet{Stein2015} proposed a method to compute the statistical significance of structure (e.g., a possible jet emitting from a quasar) in the fitted image. Here, we summarize the statistical model used by \citet{Stein2015}, which they term LIRA (Low-count Image Reconstruction and Analysis). Our primary goal is to provide a novel post hoc analysis of LIRA output designed to identify a cohesive region in the image corresponding to extended emission.

\begin{table*}[t]
    \centering
    \caption{Description of notation used in the models described in Section \ref{sec:ising:method}.}
    \label{tab:param}
    \begin{tabular}{ll}
    \hline
        $Y, y_i$ & Observed image ($Y$) containing photon counts ($y_i$) in pixel $i$ \\
        $\Lambda\base, \lambda_{{\rm b} i}$ & Normalized image of \emph{known} baseline component ($\Lambda\base$) with pixel-wise values $\lambda_{{\rm b} i}$ summing to one\\
        $\Lambda, \lambda_i$ & Normalized image of \emph{fitted} added component ($\Lambda$), with pixel-wise values $\lambda_i$ summing to one\\
        $\xi\base, \xi$ & Expected total count from baseline ($\xi\base$) and added ($\xi$) components\\
        $\tilde\Lambda, \tilde\lambda_i$ & Unnormalized image of added component ($\tilde\Lambda =\xi\Lambda$) with pixel-wise values $\tilde\lambda_i=\xi\lambda_i$ equaling expected\\
        \phantom{cat} & photon counts\\  
        $\Psi, \psi_i$ & Square root of unnormalized added component ($\Psi$) with pixel-wise values $\psi_i = \sqrt{\tilde\lambda_i}=\sqrt{\xi\lambda_i}$ equaling\\
        \phantom{cat} & the square root of expected photon counts (This notation is only used in the Appendix~\ref{app:gibbs}.)\\
       $\LamMC, \tilde\lambda_i^{(t)}$ & Single draw from the posterior distribution given by LIRA of $\tilde\Lambda$ with individual pixel values $\tilde \lambda_i^{(t)}$\\ 
         $Z,z_i$ & Image of pixel assignments ($Z$) with $z_i = +1$ (or $-1$) if pixel $i$ corresponds to source (or background)\\
        $\ZMC,z_i^{(t)}$ & Single draw from the posterior distribution of the image $Z$ with   individual pixel values $z_i^{(t)}$\\
        $\tau_0,\tau_1$ & Mean of the distribution of the square-root source emission ($\tau_1$) and background emission ($\tau_0$)\\
        $\sigma_0^2,\sigma_1^2$ & Variance of the distribution of square-root source emission ($\sigma_1^2$) and background emission ($\sigma_0^2)$ \\
        $\theta, \thetaMC$ & Collection of parameters, $\theta = (\tau_0,\tau_1,\sigma_0^2,\sigma_1^2) $ and a single draw ($\thetaMC$) of its values from the posterior\\ \phantom{cat} & distribution\\
        $\beta$ & Parameter of Ising distribution used in cohesion model, $\beta$ controls cohesion among pixel assignments\\
        $\betaMC$ & Single draw of the parameter $\beta$ from the posterior distribution\\
        $p\lira(\cdot)$ & Prior and posterior distributions associated with LIRA in Step~1 of three-step procedure.\\
        $p\coh(\cdot)$ & Prior and posterior distributions associated with cohesion models in Step~2 of three-step procedure.\\
    \hline
    \end{tabular}
\end{table*}

We consider an image composed of a grid of $n$ pixels with photon count $y_i$ observed in pixel $i$ and denote the full image by $Y= (y_1, \ldots, y_n)$. The image is modeled as a superposition of two Poisson processes; the first represents known or presumed features (e.g., background or known point sources), and the second represents an unknown structure. These two components are referred to as the \emph{baseline component} and the {\it added component}, respectively. We write the intensity (i.e., the expected photon counts) of the baseline and the added components as $\xi\base\Lambda\base$ and $\xi\Lambda$, respectively, where $\xi\base$ and $\xi$ are scale parameters representing the total expected count of each of the two components, $\Lambda\base = \{\lambda_{{\rm b}i}, i=1,\dots,n\}$ and
$\Lambda = \{\lambda_i, i=1,\dots,n\}$, with the components of both $\Lambda\base$ and $\Lambda$ summing to one. Thus, $\Lambda\base$ and $\Lambda$ are each probability vectors that represent the distribution of photons across the image in the baseline and added components, respectively. While the probability vector notation of $\Lambda\base$ and $\Lambda$ is more natural for the mathematical formulation of LIRA, a boundary for an extended source is more easily described in terms of its expected pixel counts. Thus,
we sometimes write $\tilde\Lambda = \xi\Lambda$ to represent the expected pixel counts of the added component, with $\tilde\Lambda = \{\tilde\lambda_i =\xi\lambda_i, i=1,\dots,n\}$. 
(See Table~\ref{tab:param} for a summary of our notation.) Formally, we model the observed photon counts, $Y$, with a Poisson distribution with mean equal to a linear combination of the unknown $\Lambda$ and the known $\Lambda\base$,
\begin{equation}
\label{eq:ising:lira}
    y_i | \Lambda, \xi, \xi\base \sim \mbox{Poisson}\left( \sum_{j=1}^n P_{ij}A_j ( \xi\lambda_{j} +  \xi\base \lambda_{{\rm b}j})\right) \; ,
\end{equation}
where $P_{ij}$ represents the point spread function (PSF) and equals the probability that a photon originating in pixel $j$ is recorded in pixel $i$ and $A_j$ represents the exposure map or  efficiency of photon detection at pixel $j$. Although the PSF and the exposure map must be estimated in practice, we treat both as fully known. When fitting model Equation~\ref{eq:ising:lira}, $\Lambda$, $\xi$, and $\xi\base$ are treated as unknown parameters, with $\Lambda$ being of primary interest when mapping or estimating unknown structure in an astrophysical source. The baseline component, $\Lambda\base$, is designed to capture known structure (e.g, background and point sources), and thus is not fit, but its scale is fit to the data via $\xi\base$.\footnote{\citet{Stein2015} discusses possible uncertainty in $\Lambda\base$, e.g., in terms of the location of point sources. For simplicity, we assume $\Lambda\base$ is completely known and refer readers interested in the more general formulation to  \citet{Stein2015}.}

LIRA requires prior distributions for $\Lambda$, $\xi$ and $\xi\base$ and uses a multiscale smoothing prior distribution, $p\lira(\Lambda)$, for $\Lambda$ \citep{Esch2004,Connors2007,Stein2015, nowa:kola:00}. This prior provides a flexible class of models while ensuring stability in the fit.  
We start by dividing a $2^d\times 2^d$ image into its four quadrants and reparameterizing $\Lambda$ into the proportion of counts expected in each quadrant. 
These proportions are smoothed using a Dirichlet prior distribution. The image quadrants are hierarchically split into nested subquadrants and the relative expected counts at each level 
are similarly smoothed via a hierarchy of Dirichlet priors. The smoothing parameters at each level of the hierarchy are allowed to differ, thus allowing different degrees of smoothing at different levels of resolution. In this way, we might expect little smoothing at the lowest level of resolution and more smoothing at higher levels. The parameters controlling the smoothing are themselves assigned a hyper prior distribution, allowing the degree of smoothing at each level of the hierarchy to be fit to the data. The full mathematical details of this multiscale prior and a description of how cycle spinning is used to avoid artifacts associated with quadrants and subquadrants can be found in \citet[Section~2.3]{Stein2015}. Using this specification of the added component, we focus on images that are cropped to $2^d\times 2^d$ pixels for some integer $d$.

The parameters $\xi$ and $\xi\base$ play different roles in the model and thus have different prior distributions. Because $\xi$ specifies the total expected count of the added component, its prior distribution, $p\lira(\xi)$, must be flexible and allow both for values near zero if the baseline model is adequate and large values if the added component is significant. Following \citet{Stein2015} we use a skewed Gamma distribution with mean and variance equal to 20; this ensures appreciable prior probability near zero with probability also extending to large values. The parameter, $\xi\base$, on the other hand, specifies the total expected count associated with the baseline component. Because the data typically provide information to constrain $\xi\base$, we use a diffuse prior distribution, specifically,  $p\lira(\xi\base) \propto \xi\base^{0.001 - 1}$.

Inference under Equation~\ref{eq:ising:lira} is based on the Bayesian posterior distribution, 
\begin{equation}
\label{eq:LIRA-post}
  p\lira(\Lambda, \xi, \xi\base \mid Y) \propto p(Y | \Lambda, \xi, \xi\base) \,p\lira(\Lambda) \, p\lira(\xi) \, p\lira(\xi\base), 
\end{equation}
where $p(Y | \Lambda, \xi, \xi\base)$ is given in Equation~\ref{eq:ising:lira} and $p\lira(\Lambda)$, $p\lira(\xi)$, and $p\lira(\xi\base)$ are the (independent) prior distributions used by LIRA for $\Lambda$, $\xi$, and $\xi\base$ (described in the previous two paragraphs). LIRA provides an MCMC sample from the posterior distribution in Equation~\ref{eq:LIRA-post}. In practice, the initial MCMC iterations are treated as ``burn-in'' and removed from the sample to allow the MCMC sampler to converge to its target posterior distribution \citep[e.g.,][]{sten:etal:18ii}. We denote the post burn-in sample of $\tilde\Lambda$ by $\{\LamMC=\Lambda^{(t)}\xi^{(t)}, \hbox{for } t=1,\ldots,T\lira\}$, where $t$ indexes the MCMC sample of size $T\lira$.
(We use the expected count notation $\tilde\Lambda$ rather than the probability vector notation $\Lambda$ and $\xi$ in the rest of this paper.) We estimate $\tilde\Lambda$ with its posterior mean, obtained numerically by averaging over the MCMC sample
\begin{equation}
\hatLamtilde =   {1\over T\lira}\sum_{t=1}^{T\lira}\LamMC.  
\label{eq:hat.lam.tilde}
\end{equation}

\subsection{Motivating Example: An Extragalactic Jet}
\label{sec:jet}

Supermassive black holes generate relativistic jets extending out to hundreds kiloparsec distances from their origin. These jets deposit their energy into intergalactic medium and contribute to the feedback processes driving the evolution of structures in the universe. X-rays provide unique constraints on the jets' physical parameters and radiation processes. 
In the past two decades, the high angular resolution and sensitivity of \chandra\ enable studies of kpc-scale morphology of resolved X-ray jets \citep[e.g.,][]{Harris2006,Siemiginowska2007,Hardcastle2016,Reddy2021,Reddy2023}. However, X-ray jets are faint and separating jet features, such as knots or hot spots, from a diffuse continuous jet or ISM emission has been challenging. On the other hand, the scale and size of these features provide important data for understanding the physics of relativistic jets and their impact on the environment.

\begin{figure*}
    \begin{center}
    \begin{tabular}{cc}
         %,trim={2.5cm 2cm 1.5cm 4cm},clip
         \includegraphics[width=0.4\textwidth]{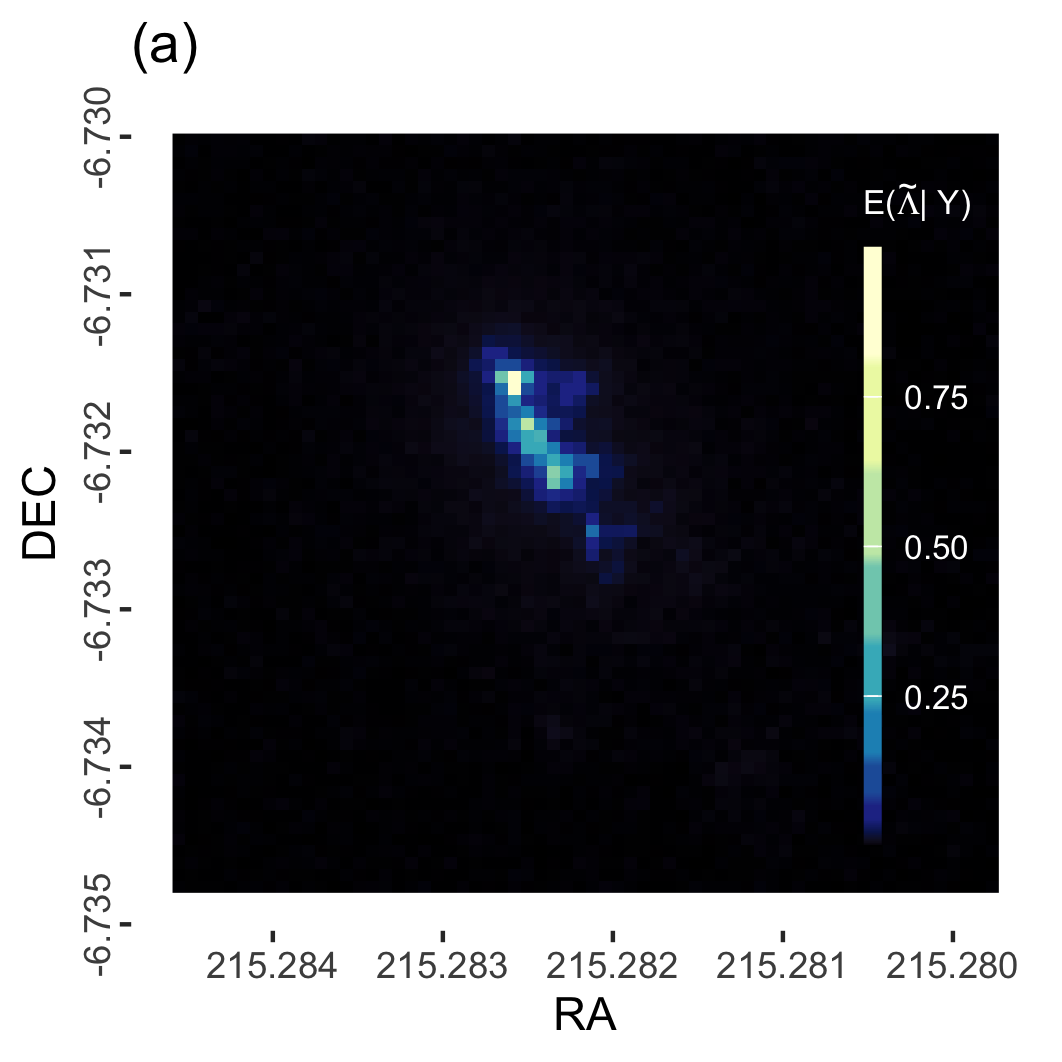} 
         & 
         \includegraphics[width=0.4\textwidth]{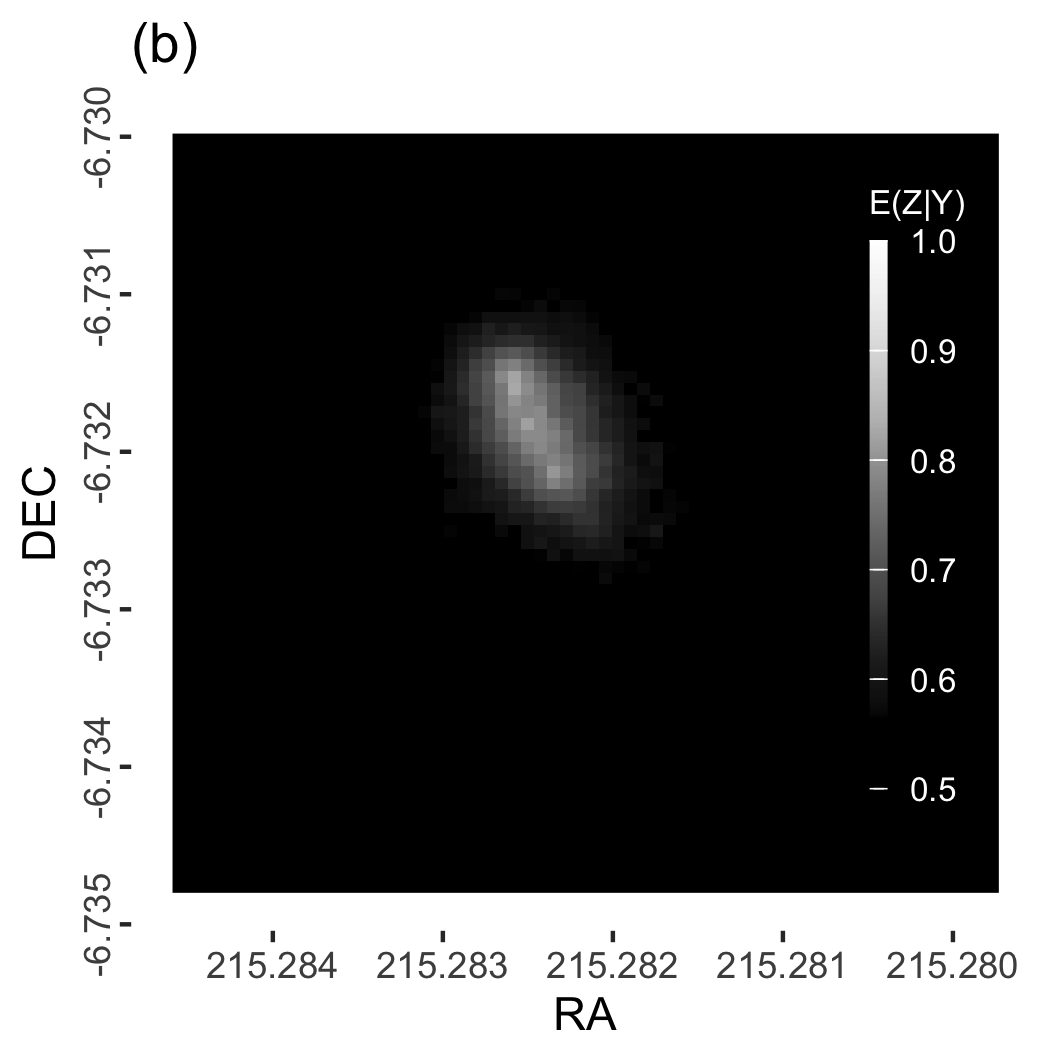}
         \\
          \includegraphics[width=0.4\textwidth]{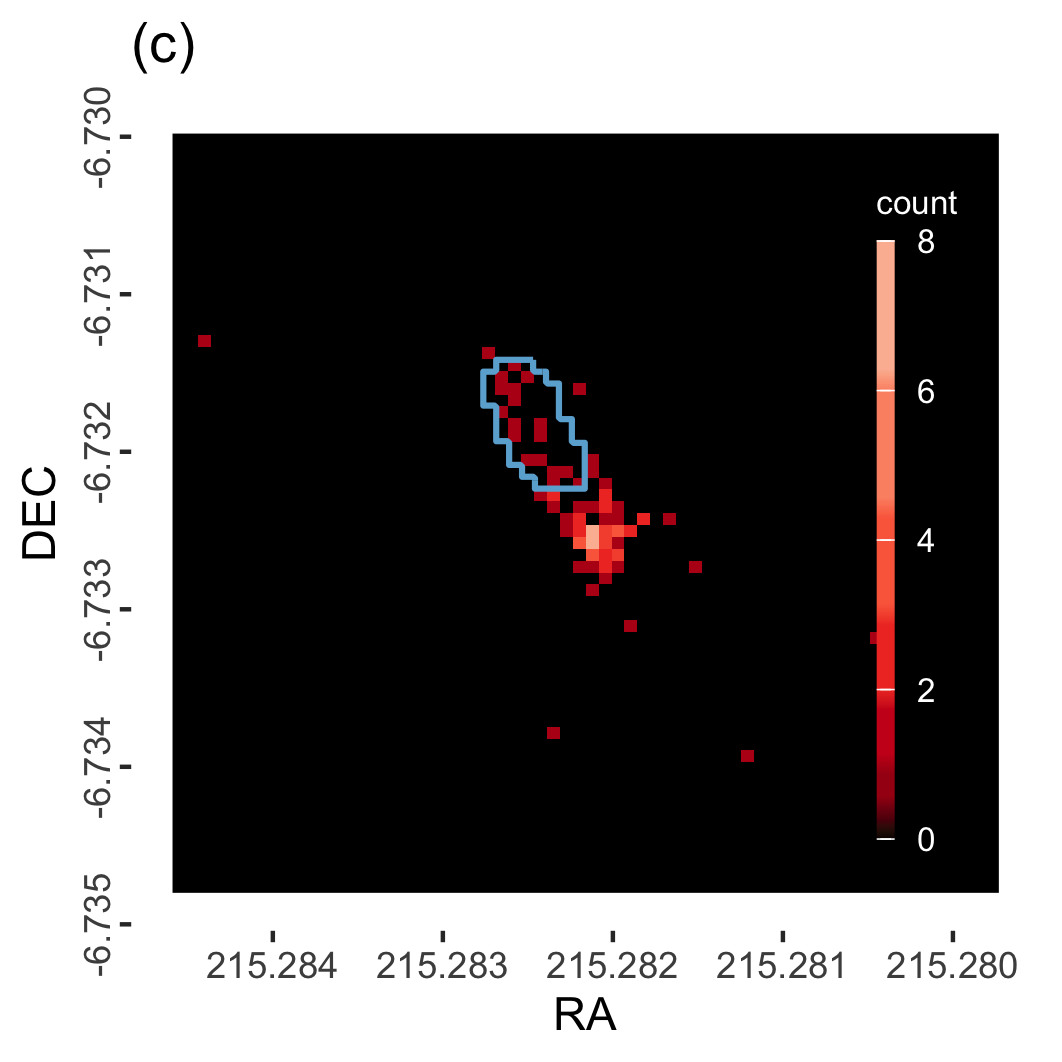}
          &
         \includegraphics[width=0.4\textwidth]{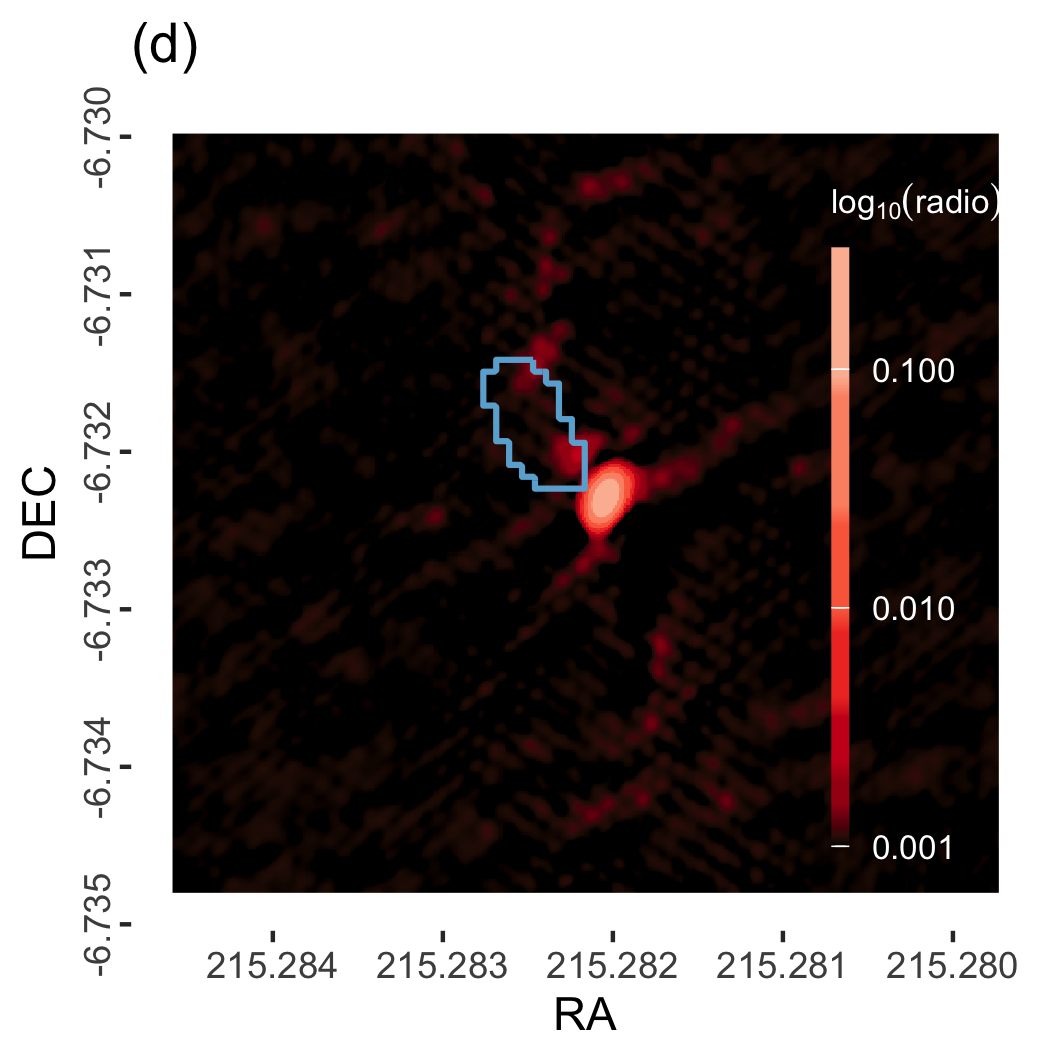}
    \end{tabular}
    \end{center}
    \caption{
    {Estimating the boundary of a jet in the} 
    \chandra\ observation of the $z = 3.69$ quasar PKS~J1421-0643 (ObsID~7873). 
    (a) $\hatLamtilde$, the posterior mean of the LIRA added component, i.e., the estimated expected deconvolved counts attributed to the jet, excluding the central galaxy. 
    (b) $\hatZ$, the map of the probability that each pixel is association with the extended (jet) source. 
    (c) The best-fit boundary of the jet overlaid on the original \chandra\ X-ray image. 
    (d) The best-fit boundary of the jet overlaid on a radio image of PKS~J1421-0643 (VLA 4.9\,GHz; from \citealt{McKeough2016}) (As the radio image can be zero or negative, in panel (d) we plot $\log_{10}(\max(\hbox{radio},0) + 0.001)$.) The color scale shows the radio measurement units in Jy/beam, beam size of $0.\arcsec66 \times 0.\arcsec38$ at PA=$-20.^o7$.} 
    \label{fig:ising:7873}
\end{figure*}

Jets associated with high redshift quasars are best suited for testing X-ray radiation processes, but they are faint and relatively short in length \cite[e.g.,][]{Harris2006,Cheung2008}. Therefore, \chandra\ observations have mostly focused on low-redshift ($z<1$) quasars, and only a few high-redshift quasar jets ($z>3.5$) have been studied to date \citep{McKeough2016,worr:etal:20,Snios2022}.
Among these is the $z = 3.69$ quasar PKS J1421-0643, which was first observed by \chandra\ on 2007 June 04 (ObsID 7873, 3.3\,ks, Figure~\ref{fig:ising:7873}(c)).  We use ObsID 7873 as an illustrative running example of our proposed method, using it to estimate the boundary of the jet. 
These data have previously been modeled with LIRA \citep{McKeough2016} with the radio image defining the jet regions to evaluate the statistical evidence for the jet \citep[as in][]{Stein2015}.
Following \citet{McKeough2016}, we treat the core of the quasar as a point source, which together with a uniform background makes up $\Lambda\base$ in Model~\ref{eq:ising:lira}. The jet is modeled using an added component $\Lambda$ and we develop and apply a new procedure to estimate the extent of the jet with quantified uncertainty. Figure~\ref{fig:ising:7873}(a) displays a heat map of $\hatLamtilde$, the posterior mean of $\tilde\Lambda$, obtained with LIRA. Because the core is included in the baseline component, $\tilde\Lambda$ can be interpreted as the expected deconvolved count of photons associated with the jet. As discussed in \citet{McKeough2016}, LIRA elicits details of the jet at higher resolutions than could otherwise be obtained.  We return to this example, throughout Section~\ref{sec:lira} as a running illustration of our three-step procedure, showing how we model the region associated with the jet, quantify its uncertainty, and produce a best-fit boundary for the jet. In Section~\ref{sec:ising:results} we apply our three-step procedure to another high redshift quasar, 0730+257 ($z = 2.69$, ObsID 10307, 20.1\,ks) with a more complex jet structure, and discuss and compare the results.

\subsection{Adding Spatial Cohesion to the LIRA Model}
\label{sec:cohesion}

Although LIRA allows for multiscale structure within an extended source, it does not formalize a boundary for the source or identify a region in the image associated with the source. Nor does the LIRA prior quantify the expectation that an extended source be spatially cohesive. To address this, we consider an extended source to be
\begin{enumerate}
\itemsep -3pt
\item[(i)] a cohesive or connected region where 
\item[(ii)] the photon emission rate tends to be higher than background.
\end{enumerate}
In order to estimate the location and range of a cohesive extended source, we augment the LIRA model with an additional prior that is specified in terms of these two characteristics of an extended source. We refer to this prior as the \emph{cohesion} prior to distinguish it from the multiscale and other priors used by LIRA and described in Section~\ref{sec:lira}.  We refer to the overall model, LIRA augmented with the cohesion prior, as the LIRA-Ising model. 

We define an indicator variable that determines which pixels are associated with the extended source. Specifically, let $Z=(z_1,\ldots, z_n)$ with $z_i = +1$ if pixel $i$ is part of the extended source and $z_i = -1$ if it is not. Implicitly, this definition assumes we are considering a region with at most one extended source or if there are multiple extended source regions we treat them as one (perhaps discontinuous) region.

The first characteristic of an extended source is that it is spatially cohesive. That is, the values of $z_i$ in neighboring pixels tend to be equal. To quantify this, we assume an Ising model as the hyper-prior distribution on $Z$, i.e., 
\begin{equation}
    p\coh( Z\mid \beta)  = \frac{1}{Q(\beta)}  
    \exp \left( \beta \sum_{d(i,j)=1} z_{j}z_{i}\right)\; ,
    \label{eq:ising:ising}
\end{equation} 
where the subscript `Coh' stands for `cohesion' as in `cohesion prior', $d(i,j)$ is a distance metric with $d(i,j)=1$ when $i$ and $j$ are adjacent pixels and  $Q(\beta)$ is the so-called partition function and acts as a normalizing constant, i.e., $Q(\beta) = \sum_Z \exp ( \beta \sum_{d(i,j)=1} z_{j}z_{i})$. The parameter $\beta > 0$ is proportional to an {\it inverse} `temperature' meaning higher values of $\beta$ result in more `freezing', i.e.,   cohesion among the pixel assignments and lower values correspond to a noisier array of pixel assignments. 

The Ising model is linked to the cohesion of the extended sources by specifying a conditional distribution for expected pixel count for the added component, i.e., $\tilde\lambda_i=\xi\lambda_i$, given $z_i$. (Recall that $\xi$ is the total expected count from the added component and $\lambda_i$ is the expected proportion of these counts attributed to pixel $i$.)  The second characteristic of an extended source is that its expected pixelwise emission rates tend to be higher than those of the background. To formalize this, we model (approximately)  the square root\footnote{We expect the distribution of the $\tilde \lambda_i$ to be non-negative and right skewed, with many values near or equal to zero. Both the square-root and the log transformations are known to be normalizing for such distributions, i.e., either $\sqrt{\tilde\lambda_i}$ or $\log(\tilde\lambda_i)$ are more appropriately modeled as Gaussian than the untransformed $\tilde\lambda_i$. In practice, we find that the MCMC sample includes values of the $\lambda_i$ that are numerically equal to zero, precluding the use of a log transformation in Equation~\ref{eq:sqrt-lam}. In this case, we find that 
the square-root transformed distribution matches the normal distribution reasonably well. We are nonetheless cautious in over interpreting the means and variances  of the normal distributions in Equation~\ref{eq:sqrt-lam}. Rather, we view this model as a data-driven and expedient solution for classifying pixels into those associated with and not associated with the extended source.} of the components of $\tilde\Lambda$ given $Z$ as a mixture of Gaussian distributions, i.e., we specify 
$p\coh(\tilde\Lambda | Z,\theta)$ via

\begin{equation}
    \sqrt{\tilde\lambda_{i}}\mid z_i, \theta  \sim 
    \begin{cases}
        \mbox{Normal}(\tau_0, \, \sigma_0^2) & \mbox{if} \ z_i = -1 \\
        \mbox{Normal}(\tau_1, \, \sigma_1^2) & \mbox{if} \ z_i = +1
    \end{cases}
    \label{eq:sqrt-lam}
\end{equation} 

where $\theta = (\tau_0,  \tau_1, \sigma_0^2, \sigma_1^2)$,  $\tau_0$ is the mean of the square-root intensity of the background and $\tau_1$ is the mean of the square-root intensity of the extended source; $\tau_1 >\tau_0$ by definition. We postulate different variances ($\sigma^2_0$ and $\sigma^2_1$) because we expect the background pixel intensities to be less variable (and clustered closer to zero) and the intensities associated with the extended source to be more variable. The square-root transformation in Equation~\ref{eq:sqrt-lam} is motivated by the expectation that the (non-negative) distribution of pixel intensities is right skewed with many values equal to zero or very near zero. 
%(In practice, individual $\lambda_i$ can be so near zero that $\log(\lambda_i)$ is numerically unstable in our fitting algorithms, see Section~\ref{sec:ising:mcmc}.)

To complete our specification of the cohesion prior, we apply Bayes Theorem using Equations~\ref{eq:ising:ising} and~\ref{eq:sqrt-lam} to derive
\begin{equation}
\label{eq:coh-prior}
    p\coh(Z, \theta, \beta \mid \tilde\Lambda) =
    \frac{p\coh(\tilde\Lambda \mid Z, \theta) p\coh(Z|\beta) p\coh(\theta,\beta)}{p\coh(\tilde\Lambda)},
\end{equation}
where $p\coh(\theta, \beta)$ is the hyper-prior distribution for $\theta$ and $\beta$ and $p\coh(\tilde\Lambda)$ is the marginal distribution of $\tilde\Lambda$ implied by the overall cohesion prior. Since the cohesion prior only enters the LIRA-Ising model through the conditional distribution given $\tilde\Lambda$ in Equation~\ref{eq:coh-prior}, inference for $\tilde\Lambda$ is unaffected; see Section~\ref{sec:fitting}.

To simplify computation, we specify $p\coh(\theta, \beta)$ independently for $\theta$ and $\beta$, using a moderately informative conjugate prior (conditional on $Z$) for $\theta$. In our numerical studies, we assume $\tau_0 | \sigma_0^2 \sim \mbox{Normal}(5,\sigma_0^2)$ with $\sigma_0^2 \sim 10/\chi^2_{(10)}$, where $\chi^2_{(10)}$ is a chi-squared random variable with 10 degrees of freedom, and use exactly the same prior for $(\tau_1,\sigma_1^2)$. We assume $\beta$ follows a Gamma distribution with mean and variance both equal to 100. We provide a more general class of prior distribuitons for both $\theta$ and $\beta$ in Appendix~\ref{app:gibbs}.

\subsection{Steps 1 and 2: Fitting the LIRA-Ising Model for Spatial Cohesion}
\label{sec:fitting}

Our strategy is to use the cohesion prior to construct the posterior distribution of $Z$, that is, the posterior distribution of the set of pixels associated with the extended source, given $\tilde\Lambda$ and then average this distribution over the posterior distribution of $\tilde\Lambda$ obtained with LIRA (given in Equation~\ref{eq:LIRA-post}). To justify this procedure, we can express the joint posterior distribution of the full set of unknown parameters as
\begin{align}
\label{eq:post-full}
\nonumber
p(\tilde\Lambda, \xi\base,Z, \theta, \beta \mid Y) 
\propto \\
\nonumber
\propto
p(Y\mid \tilde\Lambda, \xi\base, Z,\theta,\beta) \,
p\coh(Z,\theta,\beta 
\mid \tilde\Lambda)\,
&p\lira(\tilde\Lambda) \, p\lira(\xi\base)
\nonumber\\
\propto
p(Y\mid \tilde\Lambda, \xi\base) \,
\nonumber
p\lira(\tilde\Lambda) \, p\lira(\xi\base)\,
\nonumber
&p\coh(Z, \theta, \beta \mid \tilde\Lambda) 
\nonumber\\
\propto 
p\lira(\tilde\Lambda, \xi\base \mid Y)\,
p\coh(Z, \theta, \beta \mid \tilde\Lambda),
\end{align}
where the final two terms are given in Equations~\ref{eq:LIRA-post} and \ref{eq:coh-prior}, respectively. 
We use Equation~\ref{eq:post-full} as the basis of inference of the boundary. The marginal posterior distribution for $\tilde\Lambda$ under Equation~\ref{eq:post-full} is equivalent to the LIRA posterior distribution in Equation~\ref{eq:LIRA-post}. Thus, inference for $\tilde\Lambda$ is the same under the LIRA and LIRA-Ising models, see Appendix~\ref{app:joint-post} for details.
Thus, {\bf Step~1} of our three-step procedure runs LIRA as normal, for $t\lira$ post-burnin iterations.

To obtain an MCMC sample of $(Z,\theta,\beta)$ we run a Gibbs sampler with target distribution  
\begin{equation}
    p\coh(Z, \theta, \beta \mid  \LamMC) \propto 
    p\coh(\LamMC \mid Z, \theta ) \; p\coh( Z \mid \beta)\;
    p\coh(\theta, \beta), 
    \label{eq:ising:posterior}
\end{equation}
for each of the $t\lira$ values of $\LamMC\sim p\lira(\tilde\Lambda, \xi\base \mid Y)$ obtained 
in Step~1.  In {\bf Step~2}, we run a separate Gibbs sampler for each of these $t\lira$ values of $\LamMC$.
(Sampling from Equation~\ref{eq:ising:posterior} requires a novel Gibbs sampler, see Appendix~\ref{app:gibbs} for details.) Each Gibbs sampler is run long enough to obtain a post-burnin sample of size $S$ from the posterior in Equation~\ref{eq:ising:posterior} and these samples are combined to obtain an overall sample of size $T = St\lira$, denoted by $\{(\LamMC, \ZMC, \thetaMC, \betaMC), t=1,\ldots T\}$.  In this way, we account for the uncertainty in $\tilde\Lambda$ in the probabilistic classification of the image pixels. 

With the MCMC sample $\{\ZMC, t=1\ldots, T\}$ in hand, we can compute the probability map,
%\todo{\tiny XL: The same issue with the expectation notation}
\begin{equation}
\hatZ = {1\over T\lira} \sum_{t=1}^{T\lira}\ZMC.  
\label{eq:hatZ}
\end{equation}
 This is a $2^d\times 2^d$ matrix, with each element corresponding to a pixel in the image and equal to the posterior probability that this pixel is associated with the extended source.  Figure~\ref{fig:ising:7873}(b) illustrates the probability map for the high redshift quasar PKS~J1421-0643 (using the data of ObsID~7873). Similarly, the area of the extended source (in pixels) is equal to the sum of the elements of $Z$, i.e., $z_+=\sum_i z_i$. Thus, we can obtain a sample of the posterior distribution of the extended source area as $\{z_+^{(t)}, t=1,\ldots T\}$.

\subsection{Implementing Steps~1 and 2}

\label{sec:ising:mcmc}

In the numerical studies in Sections~\ref{sec:ising:validate} and \ref{sec:ising:results} we take the following approach. In Step~1, we run LIRA from 3000 iterations, discard the first 2000 iterations as burn-in, and set $t\lira =1000$. In Step 2 we run a separate Gibbs sampler for all 1000 values in the post-burn-in LIRA sample. Starting with $\LamMCone$, we ran the Gibbs sampler for 500 iterations. We then take the final draw from this initial run and use it as the starting value for each of the $t\lira=1000$ Gibbs samplers, run independently on each $\LamMC$ in the posterior sample from LIRA.  After each Gibbs samplers is run for 50 iterations, we sample a single pixel assignment $Z^{(t)}$ for each $\LamMC$ from LIRA, resulting in 1000 pixel assignment draws from the posterior. When drawing $\beta$ within the Gibbs sampler, we suggest a jump standard deviation of $0.01$, and we take a draw after a burn-in of 20 iterations. When sampling the spin states, we iterate the Swendsen-Wang algorithm and take a single sample after a burn-in of 50 iterations. (See Appendix~\ref{app:gibbs} for details.)  We use the same sampling strategy for all simulations and applications in Sections \ref{sec:ising:validate} and \ref{sec:ising:results}. It proved robust to the number of burn-in iterations, i.e., lengthening the brun-in did not alter the results.

\subsection{Step~3: Estimating the Extended Source Region Boundary under the LIRA-Ising Model}
\label{sec:ising:optimal}

Finally, in {\bf Step~3}, we estimate the boundary of the extended source with the marginal MAP estimate of $Z$, i.e., the value $\hat Z$ that maximizes $p(Z\mid Y) = \int p(Z, \theta, \beta \mid  \tilde\Lambda, Y) \; p\lira(\tilde\Lambda \mid Y) \, {\rm d}\theta \, {\rm d} \beta \, {\rm d}  \tilde\Lambda $. 
To avoid optimization of the discrete parameter $Z$, we construct a collection of values of $Z$, ${\cal Z} =\{Z_1, \ldots, Z_M\}$, which we expect to contain values at or near the maximum of $p(Z\mid Y)$. We then construct a Monte Carlo estimate of the ratio of integrals, $p(Z_1 \mid Y) / p(Z_2 \mid Y)$, to determine which of $Z_1$ or $Z_2$ has a higher marginal posterior probability. The value with the highest probability is then compared with $Z_3$. Cycling through $\cal Z$ this way, we set $\hat Z$ to the value of $Z$ in $\cal Z$ that maximizes $p(Z\mid Y).$ (The ratio is more stable numerically than direct Monte Carlo estimation of $p(Z\mid Y)$, see Appendix~\ref{app:Zmargin} for details.) 

To construct $\cal Z$, we define a {\it neighborhood statistic} that can be computed for each pixel.
For a pixel associated with the extended source (i.e., a pixel $i$ with $z_i=+1$), 
the neighborhood statistic is the proportion of pixels adjacent to pixel $i$ that are also associated with the extended source. For a pixel not associated with the extended source (i.e., a pixel $i$ with $z_i = -1$), the neighborhood statistics is set to zero. Mathematically, we define the neighborhood statistic to be 
\begin{equation}
    \Phi_{i}(Z) = \frac{\sum_{j \in \{d(i,j)=1\}} (z_i +1)(z_j + 1)}  {4\sum_{j \in \{d(i,j)=1\}} 1} 
\end{equation}
and compute its posterior mean for each pixel, i.e., $\bar \Phi_{i}= \sum_t^T \Phi_i(\ZMC)$. (Note that $(z_i+1)/2$ equals one (zero) for pixels associated (not associated) with the extended source.) The first element of $\cal Z$ attributes only one pixel to the extended source: the pixel with the largest mean neighborhood statistic. The second element attributes only the two pixels with the largest mean neighborhood statistic to the extended source. We continue in this way, adding a single pixel to the extended source in each subsequent element of $\cal Z$ in order of the mean neighborhood statistic until we have an element of $\cal Z$ that attributes all the pixels to the extended source. Finally, we add the Monte Carlo sample $\{\ZMC, t=1, \ldots, T\}$ to $\cal Z$ since they represent the dense space of the posterior distribution and thus are suitable candidates for $\cal Z$. 

The resulting best-fit boundary for QSO\,0730+257 (ObsID~7873) is plotted in Figures~\ref{fig:ising:7873}(c) and~\ref{fig:ising:7873}(d) (see Section~\ref{sec:ising:results}). Figure~\ref{fig:ising:7873}(c) compares the boundary with the raw \chandra\ X-ray counts and Figure~\ref{fig:ising:7873}(d) compares it with a VLA radio image of the same source (see Table~2 of \citet{McKeough2016}).

\section{Validation}
\label{sec:ising:validate}

We present two simulation studies that are designed to validate the LIRA-Ising model and our three-step procedure. In the first simulation the expected counts from the extended source are {generated from} a two-dimensional Gaussian distribution; in the second simulation the expected counts follow a two-dimensional step function. Both simulations are repeated with differing levels of a uniform background. The extended source has a soft boundary in the first simulation and a hard boundary in the second. A soft boundary is more realistic for most astrophysical images, whereas the second enables us to quantify errors in the classification of pixels as being part of or not part of the extended source. (With a soft boundary, there is no ground truth for separating boundary between the extended source and the background region.)

\subsection{Simulation with a Soft Boundary}
\label{sec:soft}
\begin{figure*}
    \centering
    \includegraphics[width=\textwidth]{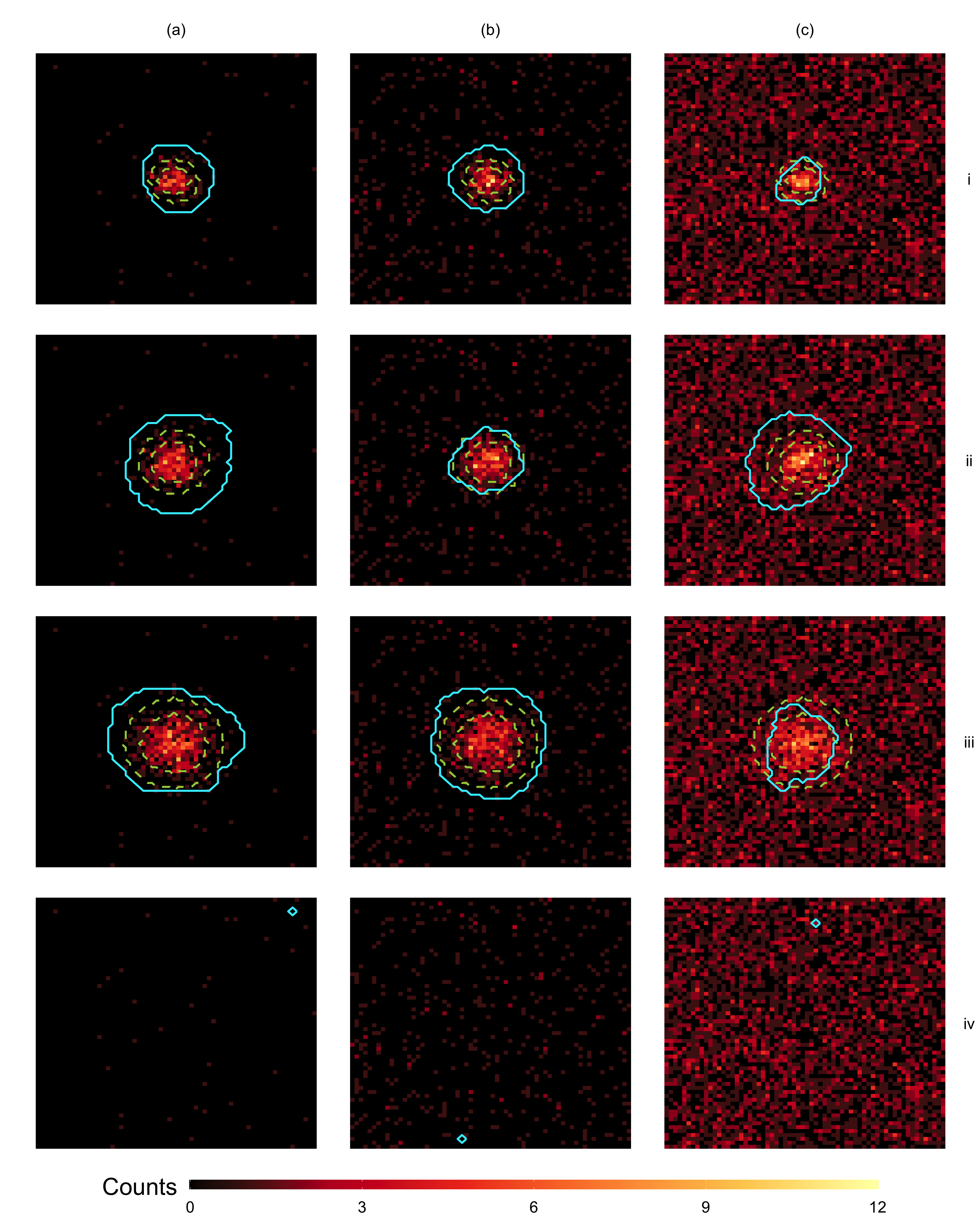}
    \caption{Simulation study with a soft boundary. Rows correspond to the the four extended source settings (from top to bottom: Gaussian with variance equal to 4, 8, and 16 pixels, and no extended source). Columns correspond to noise level (from left to right 0.01, 0.1 and 1 expected counts per pixel). Each panels shows the simulated counts and their fitted boundaries. The fitted boundaries are plotted as solid cyan curves and are compared with the $2\sigma$ and $3\sigma$ dashed green contours of the Gaussian distributions used to simulate the extended sources.}
    \label{fig:B104_bound}
\end{figure*}

We simulate a set of extended sources with soft boundaries, each in a $64\times64$ pixel image. The contours of the expected source count follow two-dimensional circular Gaussians with radius set to one of three values corresponding to the variance of the Gaussian being fixed at 4, 8, or 16 pixels. The uniform background level is also varied with expected brightness set to 0.01, 0.1 and 1 counts per pixel. An additional setting is run with no extended source (for each of the three background levels). The resulting $4\times 3$ design is illustrated in  Figure~\ref{fig:B104_bound} where rows correspond to the width of the extended source (increasing from top to bottom, with the no-extended-source setting in row four) and columns correspond to the background level (increasing from left to right). All nine source simulations have a peak intensity of five expected photons per pixel. Poisson counts are generated for each of the two-dimensional Gaussian and background templates. For consistency, the same background realizations are used across the four extended source settings. Finally the realizations are convolved with a two-dimensional Gaussian PSF with a two-pixel standard deviation. 

We deploy the LIRA-Ising model via the three-step procedure to obtain a boundary estimate for each of the 12 simulated datasets. In Step~1 we run LIRA using the same PSF as was used to generate the data and with $\Lambda\base =0$. Thus, the background and extended source are fit together as part of the multiscale added component. Figure~\ref{fig:B104_bound} plots the posterior means of the expected counts of the added component, i.e., $\hatLamtilde$
%\todo{\tiny XL: same notation issue.} 
for each of the 12 simulations. Running Steps~2 and 3, we optimize over the posterior distribution of pixel assignments to obtain the boundary estimates. In Figure~\ref{fig:B104_bound} the boundary estimates are plotted as solid cyan curves and compared with the two and three standard deviation dashed green contours of the underlying Gaussian extended source models.

Focusing first on the upper three rows of Figure~\ref{fig:B104_bound} where the simulation models include extended sources, in columns (a) and (b), the fitted boundaries corresponds to areas that are slightly larger than the $3\sigma$ contours. In the low signal to noise setting, column (c), the fitted boundary is closer to the $2\sigma$ contour. This is not unexpected. In the noisier images the low-count regions in the tails of the extended sources are easily confused with background. There is variability, however, as illustrated in row (ii) where the fitted boundary includes more of the noisy region. Again this illustrates that boundary identification is more difficult in lower signal-to-noise regimes. Finally, in row (iv) the fitted boundaries all correspond to a single pixel. Since there is no source in these simulations, the algorithm does not find a cohesive region to bound.

\subsubsection{Simulation with a Hard Boundary}

We repeat the validation study in Section~\ref{sec:soft}, but with a different set of extended sources with sharp boundaries. In particular, we replace the Gaussian extended sources with extended sources that are uniform over a square. We consider three settings with squares having 4, 8, and 16 pixels a side. In all cases the expected source count is one event per pixel. We consider the same background levels as in Section~\ref{sec:soft} for a $3\times 3$ overall design. This simulation is again conducted with $64\times 64$ images, Poisson counts, and a PSF with a standard deviation of one pixel. Although we do not expect such hard boundaries in most astronomical images, this set up has a clearly defined ground truth for the boundary and enables us to check our algorithm against this truth. It also allows us to evaluate performance as a function of the relative source and background intensities.  

\begin{figure*}
    \centering
    \includegraphics[width=\textwidth]{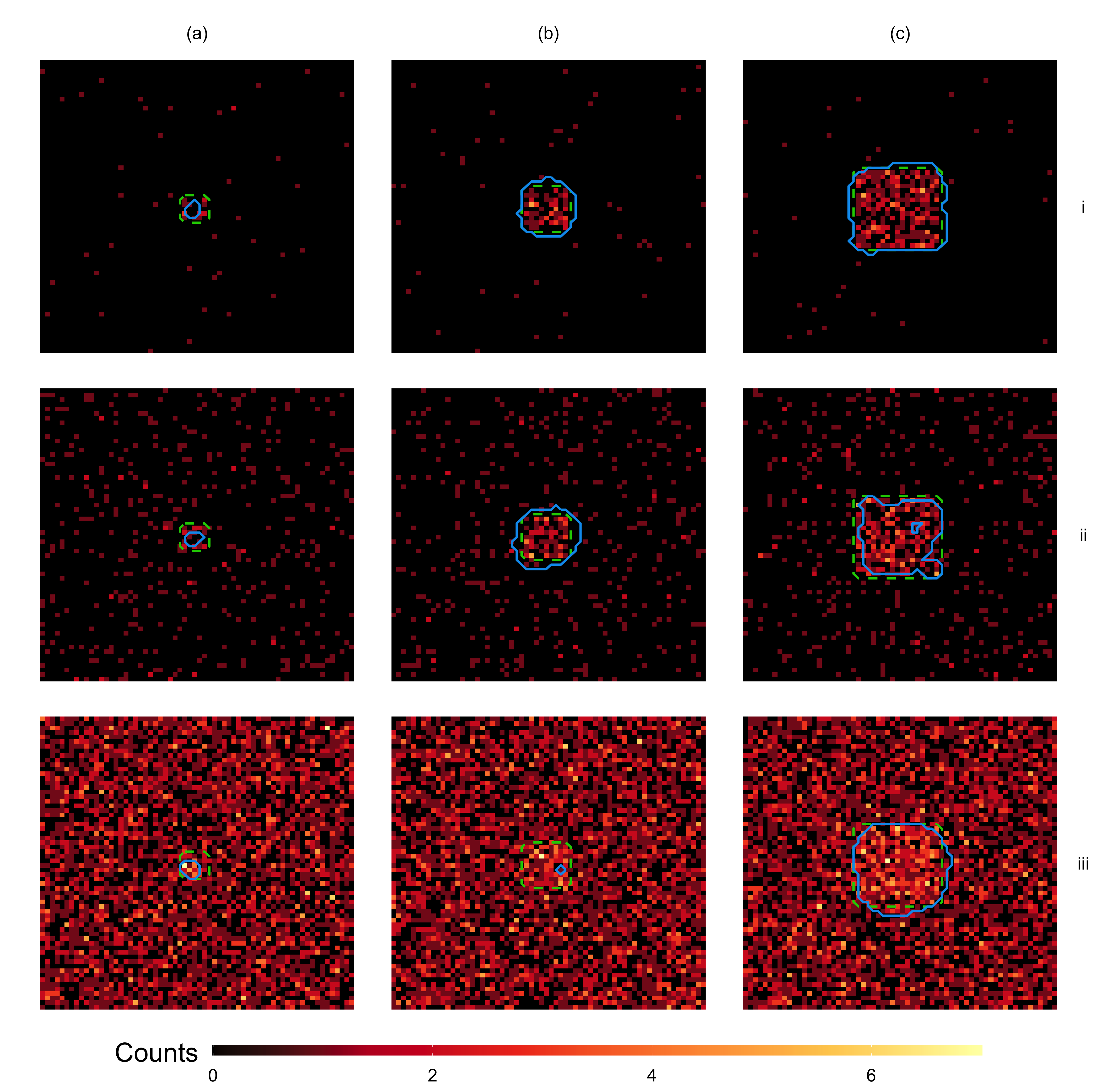}
 \caption{Simulation study with a hard boundary. Columns correspond to the three extended source settings (from left to right: squares with 4, 8, and 16 pixels to the side, centered in the field, each with a brightness of 1 count~pixel$^{-1}$). Rows correspond to noise level (from top to bottom 0.01, 0.1 and 1 expected counts per pixel). Each panel shows the simulated counts and their fitted boundaries (solid cyan). The  boundaries are compared with the the true edges of the square extended sources (dashed green).}
    \label{fig:squares}
\end{figure*}

The simulated datasets appear in Figure~\ref{fig:squares}, with columns corresponding to the size of the square extended sources (increasing from left to right) and rows corresponding to the background level (increasing from top to bottom). We apply the LIRA-Ising model via our three-step procedure as in Section~\ref{sec:soft}. The fitted boundaries are shown overlaid on the Poisson realizations in Figure~\ref{fig:squares} (blue solid line) along with the true edges of the square extended sources (green dashed line).

In all simulation settings, the extended source has only one more count per pixel on average than the background. The smallest extended sources (column (a) of Figure~\ref{fig:squares}) are partially lost in the background, but the fitted boundaries do capture at least part of the source region. This situation is particularly acute with the highest noise level where even the middle sized source is lost. The boundaries of the larger sources are much better recovered, especially in higher signal-to-noise regimes.

\section{Illustrating the method with \chandra\ data}
\label{sec:ising:results}

In this section, we fit the LIRA-Ising model and walk through our three-step procedure for the analysis of two high-redshift quasar jets. The first of these was used as a motivating example in Section~\ref{sec:lira} (PKS J1421$-$0643, Obs ID 7873, $z=3.689$, see Figure~\ref{fig:ising:7873}). The second is a quasar $0730+257$ (Obs ID 10307) at a redshift of $z=2.686$.  Both jets were first observed in radio wavelengths, where the detail of the morphology is much clearer \citep{Beasley_2002, ellison2004}. X-ray observations were performed with \chandra. \citet{McKeough2016} applied the statistical method described in \citet{Stein2015} to assess the statistical significance of the high redshift jets in pre-defined regions of interest determined from radio images. One complication with the method of Stein et al.\ and McKeough et al.\ is that in some cases, the detection power was not robust to the size and shape of the boundary determined by radio observation{; this is to be expected, since the X-ray and radio emission volumes do not necessarily coincide}. Here we take a different approach by using our three-step procedure to estimate the boundary of the region of interest, i.e., the jet. 

\begin{figure*}
    \begin{center}
    \begin{tabular}{cc}
         %,trim={2.5cm 2cm 1.5cm 4cm},clip
         \includegraphics[width=0.4\textwidth]{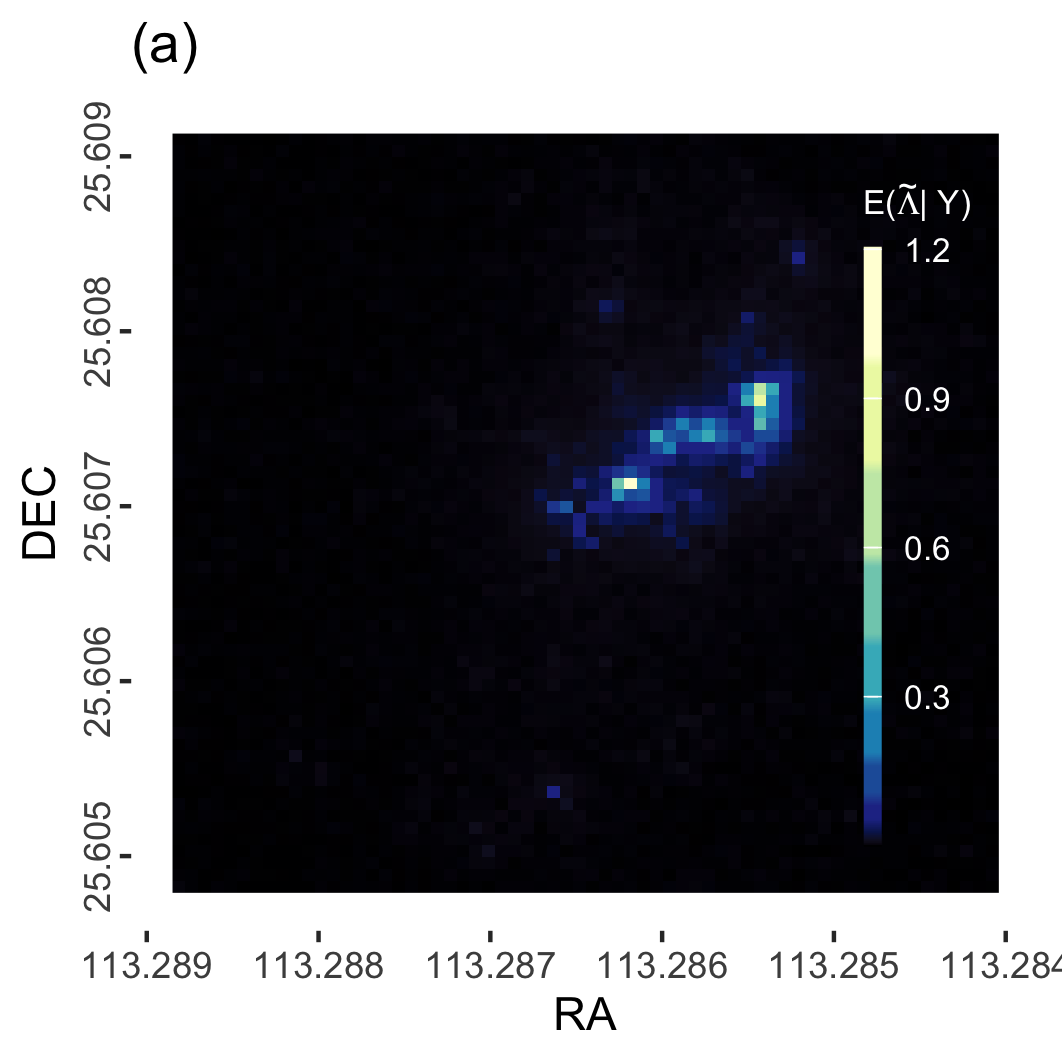} 
         &
         \includegraphics[width=0.4\textwidth]{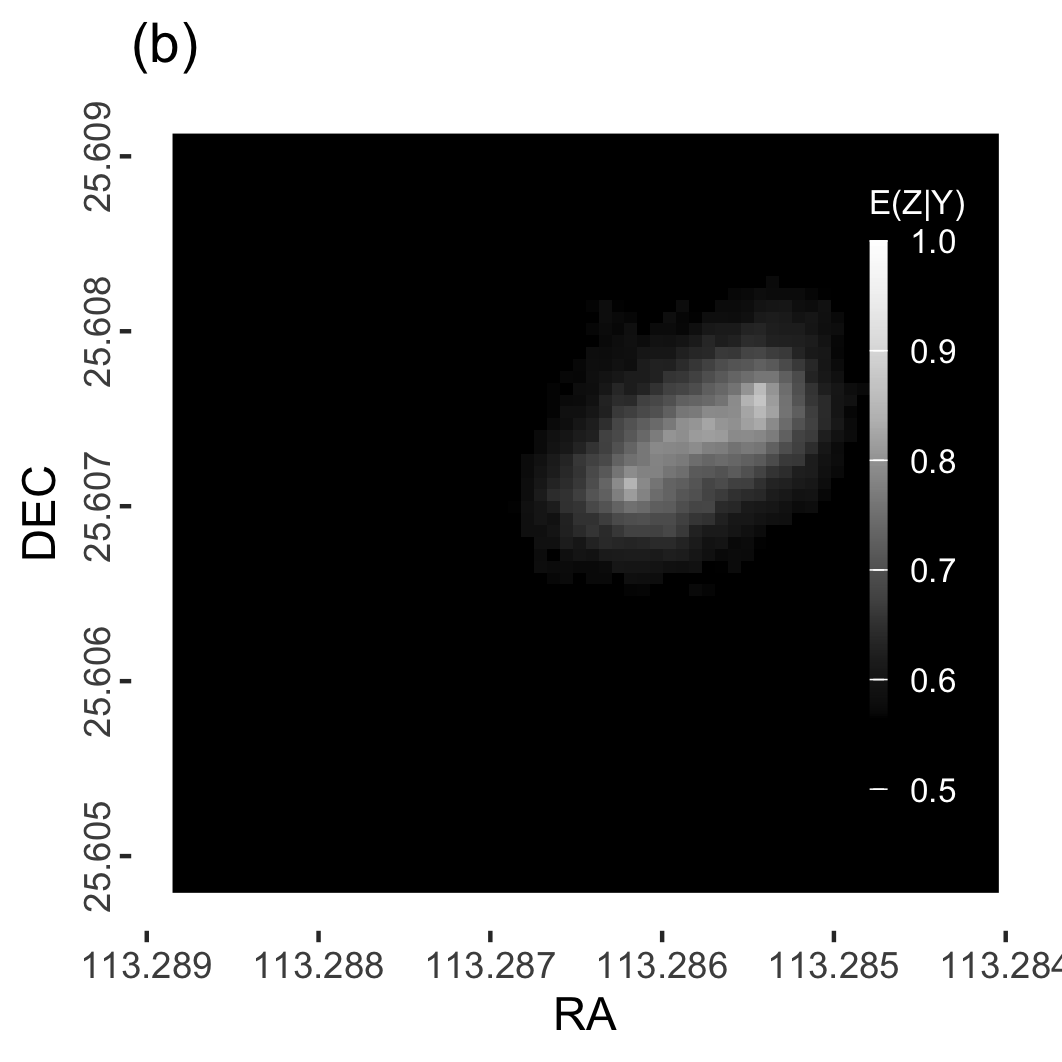}
         \\       
          \includegraphics[width=0.4\textwidth]{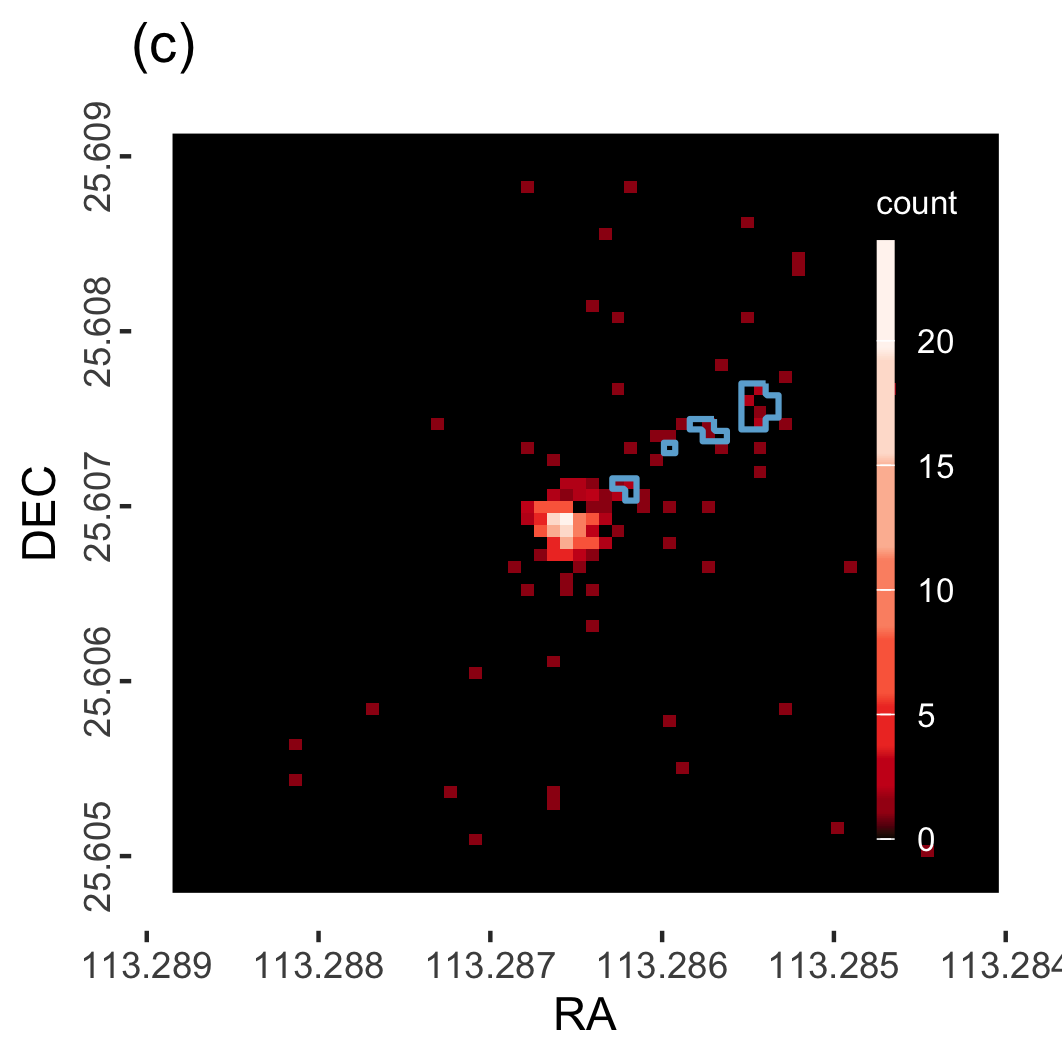}
          &
          \includegraphics[width=0.4\textwidth]{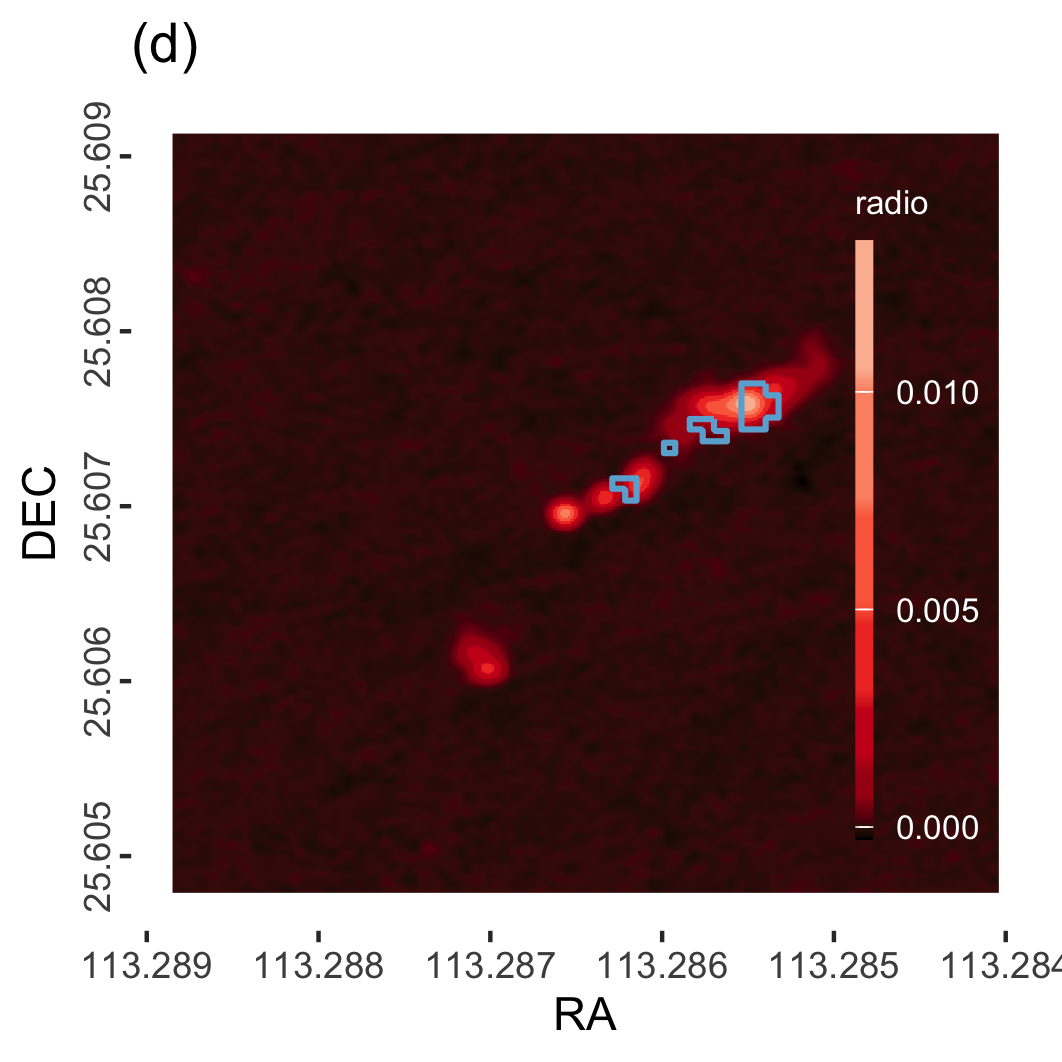}
    \end{tabular}
    \end{center}
    \caption{ 
    Identifying hot spots in a jet in the \chandra\ observation of a quasar~$0730+257$ (ObsID 10307).
       (a) $\hatLamtilde$, the posterior mean of the LIRA added component, i.e., the estimated expected deconvolved counts attributed to the jet, excluding the central galaxy. 
    (b) $\hatZ$, the map of the probability that each pixel is association with the extended (jet) source.  
    (c) The best-fit boundary of the jet overlaid on the original X-ray image. 
    (d) The best-fit boundary of the jet overlaid on a 8.7\,GHz VLA radio image of $0730+257$ from \cite{McKeough2016}. The radio measurement units are Jy/beam, with the circular beam size of 0.\arcsec35.}
    \label{fig:ising:10307}
\end{figure*}

Figures \ref{fig:ising:7873}(c) and \ref{fig:ising:10307}(c) show the raw \chandra\ X-ray images for the two quasars. It is difficult to determine the morphology of the jet from the counts alone because only a handful of high-energy photons are collected by the telescope (17 counts in PKS J1421$-$0643 jet and 12 counts in $0730+257$ jet),
particularly compared to the quasar in the center of the view. 
In Step~1 we fit both data sets with LIRA (as described in \citet{McKeough2016}). Figures \ref{fig:ising:7873}(a) and \ref{fig:ising:10307}(a) show the resulting $\hatLamtilde$, i.e., the fitted expected counts image for the two jets {with the bright central quasar modeled out}.  LIRA removes the quasar core and deconvolves the image accounting for the PSF so that the details within the jet are easier to decipher. Running Step~2 we obtain a posterior sample of the pixel assignments that we use to compute the probability maps, $\hatZ$, plotted in Figures~\ref{fig:ising:7873}(b) and \ref{fig:ising:10307}(b). These maps indicate the estimated probability that each pixel is contained within the jet. Finally, in Step~3, we optimize the posterior to obtain the boundary estimates. The estimates are overlaid on the raw X-ray counts in Figures~\ref{fig:ising:7873}(c) and \ref{fig:ising:10307}(c) and on the radio-images in Figures~\ref{fig:ising:7873}(d) and \ref{fig:ising:10307}(d) (plotted as cyan curves). In the image of PKS J1421$-$0643 (Figure~\ref{fig:ising:7873}) the three-step procedure returns a single boundary for a connected extended source. This boundary roughly resembles the boundary of the region of interest from the radio image used in \citet{McKeough2016}, but with a more form fitting shape. However, it is not uncommon in other sources to have multiple nodes and varying structure within a single jet. In the image of $0730+257$ (see Figure~\ref{fig:ising:10307}), for example, our estimated boundary algorithm indicates several disconnected regions, specifically, four distinct regions branching out from the top right of the quasar. These distinct regions highlight variations in the jet's surface brightness and indicate locations of enhanced energy dissipation and location of jet knots. The alignments and offsets between the radio and X-ray knots provide important information on physical processes in relativistic jets \cite[e.g.,][]{Reddy2021}.

\section{Conclusion}
\label{sec:ising:conclusion}

In this paper, we propose the LIRA-Ising model and develop a novel three-step Bayesian statistical procedure that deploys the model to estimate the boundary of complex, extended, irregularly shaped, diffuse astronomical structure in X-ray images. The
statistical novelty of the method lies in its expansion of the parameterization of the LIRA
multiscale method for image reconstruction
and in its layering of a second Bayesian prior distribution onto LIRA to model the intrinsic spatial cohesion of extended emission. 
In this way the overall method can be viewed as a sophisticated model-based Bayesian post-processing of the LIRA output.  Its overall prior distribution quantifies the dual and scientifically meaningful expectations of spatial cohesion of an extended source and multiscale structure in its intensity. The model is formulated in a statistically principled and computationally practical manner.
It enables us to leverage the existing LIRA method to obtain the estimated boundary with little additional computational effort. 
 We are able to quantify uncertainty in the boundary via a pixel-by-pixel posterior probability map. From an astrophysical point of view, estimating such a boundary enables more objective synthesis of the morphology of structures in astronomical images. We demonstrated this power of the method in delineating extragalactic jets.

There are a number of directions for further adaptation and generalization of the method. For example, the cohesion prior in Equations~\ref{eq:ising:ising} -- \ref{eq:sqrt-lam} could be applied directly to the Poisson counts $y_i$ rather than to their LIRA-fitted expectations $\tilde\lambda_i$, particularly in higher count data sets. This approach would involve less overhead and provide a general clustering method with a preference for spatially coherent clusters.  In this context, it is natural to consider more than two clusters by generalizing the finite mixture model in Equation~\ref{eq:sqrt-lam} to a mixture of $K$ (rather than just 2) component Gaussian distributions. Similarly, in some settings it may be useful to generalize the LIRA-Ising model by replacing Equation~\ref{eq:sqrt-lam} with a mixture of $K$ Gaussian distributions. This would allow for multiple extended structures with different intensities (per pixel) or a gradient in the intensity of a single extended structure. For example, a quasar jet might be divided into several segments of differing intensity, each of which represents one of several nodes in the jet (e.g. Figure~\ref{fig:ising:10307}). Finally, users may prefer to quantify uncertainty in the boundary with a measure of uncertainty in the location of the boundary curve itself, rather than via the posterior 
probability that each pixel is associated with the extended source. Fourier descriptors can be used to derive a very flexible parameterization of a closed boundary \citep{wang:etal:24}. Incorporating this strategy into LIRA-Ising would require a major overhaul of the model, but the parameterization would allow direct quantification of uncertainty in the location of the boundary. 

The implementation of the algorithm in R is available at \url{https://github.com/astrostat/LIRA-Ising}

    \section*{Acknowledgements}    
    This work was conducted under the auspices of the CHASC International Astrostatistics Center. CHASC is supported by NSF grants DMS-21-13615, DMS-21-13397, and DMS-21-13605; by the UK Engineering and Physical Sciences Research Council [EP/W015080/1]; and by NASA APRA Grant 80-NSSC21-K0285.    
    We thank CHASC members for many helpful discussions.
    DvD was also supported in part by a Marie-Skodowska-Curie RISE Grant (H2020-MSCA-RISE-2019-873089) provided by the European Commission.   
    AS and VK further acknowledge support from NASA contract to the Chandra X-ray Center NAS8-03060.
    BM acknowledges funding from the Engineering and Physical Sciences Research Council (EPSRC), grant number EP/S023151/1.

    Some of the computations were conducted on the Smithsonian High Performance
    Cluster (SI/HPC), Smithsonian Institution \url{https://doi.org/10.25572/SIHPC}.

    This research has made use of data obtained from the Chandra Data Archive, contained in \url{https://doi.org/10.25574/cdc.445}.This research also made use of software provided by the Chandra X-ray Center (CXC) in the application packages CIAO \citep{Fruscione2006} and {\it Sherpa} \citep{Freeman2001,Siemignowska2024}.

\appendix

\section{The Joint Posterior Distribution Under the LIRA-Ising Model}
\label{app:joint-post}

Step~1 of the three-step procedure outlined in Section~\ref{sec:fitting} samples 
\begin{equation}
    (\LamMC, \xi\base^{(t)}) \ \hbox{ from } \
    p(\tilde\Lambda, \xi_b \mid Y) 
    = 
    \frac{
    p(Y\mid \tilde \Lambda, \xi\base) \,
    p\lira(\tilde\Lambda) \, p\lira(\xi\base)
    }
  {p\lira(Y)}.
\end{equation}
Step~2 samples
\begin{equation}
   (\ZMC, \thetaMC,\betaMC) \ \hbox{ from } \
    p(Z, \theta, \beta \mid \LamMC) 
    = \frac{p\coh(\LamMC\mid Z, \theta) \,
    p\coh(Z\mid \beta) \,
    p\coh(\theta) \, p\coh(\beta)
    }{p\coh(\LamMC)
    }.
\end{equation}
Combining these two steps is mathematically equivalent to jointly sampling
\begin{eqnarray}
    \nonumber
   (\LamMC, \xi\base^{(t)},\ZMC, \thetaMC,\betaMC) \ \hbox{ from }& \
    \displaystyle\frac{
    p(Y\mid \tilde \Lambda, \xi\base) \,
    p\lira(\tilde\Lambda) \, p\lira(\xi\base)
    }{p\lira(Y)}
\times
   \frac{p\coh(\tilde\Lambda, Z, \theta,\beta)
    }{p\coh(\tilde\Lambda)
    }\\
    =& \nonumber
    \displaystyle\frac{
    p(Y\mid \tilde \Lambda, \xi\base)
    }{p\lira(Y)}
\times
   \frac{p\coh(\tilde\Lambda, Z, \theta,\beta) \,
    p\lira(\tilde\Lambda) \, p\lira(\xi\base)
    }{p\coh(\tilde\Lambda)
    }\\
    =& \label{eq:joint-post}
    \displaystyle\frac{
    p(Y\mid \tilde \Lambda, \xi\base)
    }{p\lira(Y)}
\times
   p\coh(Z, \theta,\beta \mid \tilde\Lambda) \,
    p\lira(\tilde\Lambda) \, p\lira(\xi\base),   
\end{eqnarray}
where the term after the $\times$-sign in Equation~\ref{eq:joint-post} is the joint prior distribution, with marginal prior distribution for $(\tilde\Lambda, \xi\base)$ equal to $p\lira(\tilde\Lambda) p\lira(\xi\base)$. Thus, the overall procedure relies on a coherent fully Bayesian posterior distribution that preserves LIRA's posterior distribution for $\tilde\Lambda$ (and $\xi\base$).

It is instructive to examine the conditional prior distribution, $p\coh(Z \mid \tilde\Lambda, \theta,\beta)$; note that the conditioning of $Z$ given $\tilde\Lambda$ is reversed here vis-\`a-vis Equation~\ref{eq:sqrt-lam}. To simplify our derivations, we write $\psi_i = \sqrt{\tilde\lambda_i}$ and $\Psi=\{\psi_i, i=1,\ldots, n\}.$ 
Ignoring constants that do not depend on \(Z\), its conditional prior distribuiton can be expressed as
\begin{equation}  \label{eq:log.post.z}
\log p\left(z \mid \Psi, \beta, \theta\right)
=\;
\beta \sum_{i<j} z_i z_j
\;+\;
\sum_{i=1}^n 
\phi\left(\frac{\psi_i-\tau_{(z_i+1)/2}}{\sigma_{(z_i+1)/2}}\right)
\;+\;
\text{const},
\end{equation}
where $\phi$ is the probability density function of a standard normal random variable, $\tau_{(z_i+1)/2}=\tau_{0}$ when $z_i =-1$, and $\tau_{(z_i+1)/2}=\tau_{1}$ when $z_i =1$ (likewise for $\sigma_{(z_i+1)/2}$); see Equation~\ref{eq:sqrt-lam}.
To simplify this expression, we note that for any function $f$ evaluated at $+1$ or $-1$,
\[
f(z)=\tfrac12\!\Big(f(+1)+f(-1)\Big)\;+\;\tfrac12\!\Big(f(+1)-f(-1)\Big)\,z.
\]
Applying this to the second sum in Equation~\ref{eq:log.post.z} yields
\begin{equation}
\label{eq:log.post.z.simp}  
\log p(Z\mid \Psi,\beta,\theta)
= \beta\sum_{i<j}z_i z_j\;+\;\sum_{i=1}^n h_i\,z_i,
\;+\;\text{const},
\end{equation}
where\footnote{
The factor of \(1/4\)  arises because writing
\(f(z_i)=\alpha_i+\beta_i z_i\) yields
\(\beta_i=\tfrac12\big(f(+1)-f(-1)\big)\); combining this with the
\(-\tfrac12\) factor from the Gaussian log-density gives the overall \(1/4\).
}
\begin{eqnarray} \nonumber
h_i \equiv h(\psi_i) \;&=& \;
\frac12\;\left(
\log\phi\left(\frac{\psi_i-\tau_1}{\sigma_1}\right)
-
\log\phi\left(\frac{\psi_i-\tau_0}{\sigma_0}\right)
\right)\\ \nonumber
\\ \label{eq:h.closed}
&=&\; 
\frac14\;\log\!\frac{\sigma_0^2}{\sigma_1^2}
\;+\;
\frac14\;\left(
\frac{(\psi_i-\tau_0)^2}{\sigma_0^2}
-\frac{(\psi_i-\tau_1)^2}{\sigma_1^2}
\right).
\end{eqnarray}
Thus, the distribution of $Z$ given $\Psi=\tilde\Lambda$ is an Ising model \emph{with an external (magnetic) field} governed by $\Psi$ as determined by the second sum in Equation~\ref{eq:log.post.z.simp}.

Expanding the quadratic terms in Equation~\ref{eq:h.closed}, we can rewrite
\[
h_i \;=\; A_2\,\psi_i^2 + A_1\,\psi_i + A_0,
\ \ \hbox{where} \ 
\begin{cases}
A_2 &= \tfrac14\!\left(\frac1{\sigma_2^2}-\frac1{\sigma_1^2}\right),\\[2pt]
A_1 &= \tfrac12\!\left(\frac{\tau_1}{\sigma_1^2}-\frac{\tau_2}{\sigma_2^2}\right),\\[2pt]
A_0 &= \tfrac14\!\left(
\log\!\frac{\sigma_2^2}{\sigma_1^2}
+\frac{\tau_2^2}{\sigma_2^2}-\frac{\tau_1^2}{\sigma_1^2}
\right).
\end{cases}
\]
In the special case where
\(\sigma_1^2=\sigma_0^2=\sigma^2\), $A_2=0$ and \(h_i\) become affine in \(\psi_i\), i.e., 
\[
h_i \;=\; \frac{\tau_1-\tau_0}{2\sigma^2}\,\psi_i \;+\; \frac{\tau_0^2-\tau_1^2}{4\sigma^2}.
\]

Finally, the conditional model can be linked to logistic regression in that the log odds for a single site, $z_i$, given all other sites is
\begin{eqnarray} \nonumber
\log\frac{\Pr(z_i=+1\mid Z_{-i},\Psi,\beta,\theta)}{\Pr(z_i=-1\mid Z_{-i},\Psi,\beta,\theta)}
&=& \, 2\,\!\left[\beta\sum_{j\ne i} z_j + h(\psi_i)\right]\\ \nonumber
\\
&=& \; 2\,\beta\sum_{j\ne i} z_j
\;+\; 2A_2\,\psi_i^2 \;+\; 2A_1\,\psi_i \;+\; 2A_0.\nonumber
\end{eqnarray}
Thus, the logistic predictor is quadratic in \(\psi_i\) %whenever \(\sigma_1^2\neq \sigma_0^2\) 
and the conditional prior model for $Z$ corresponds to quadratic discriminant analysis for each individual pixel with correlations determined by the Ising model. In the case of equal variances \(\sigma_1^2=\sigma_0^2\), (where \(A_2=0\)) the predictor is reduced to a linear form, corresponding to linear discriminant analysis for each individual pixel.

\section{Gibbs Sampler for the Probabilistic Classification of Pixels}
\label{app:gibbs}

Letting $\{\PsiMC, t=1,\ldots T\lira\}$ be the MCMC sample of $\Psi$, here we describe the Gibbs Sampler we use in Step~2 to obtain a Monte Carlo sample from $p(Z, \theta, \beta \mid \PsiMC, Y)$, see Equation~\ref{eq:ising:posterior}. This involves iteratively sampling the following distributions for $s=1,\ldots, S$:
\begin{enumerate}
\item $\beta\cur \sim p\coh\left(\beta \mid Z\prev, \theta\prev, \PsiMC, Y\right) = p\coh\left(\beta \mid Z\prev\right)$ 
\item $\theta\cur \sim p\coh\left(\theta \mid Z\prev, \beta\cur, \PsiMC, Y\right) = p\coh \left( \theta \mid Z\prev, \PsiMC \right)$
\item $Z\cur \sim p\coh\left(Z \mid \theta\cur, \beta\cur, \PsiMC, Y\right)= p\coh\left(Z \mid \theta\cur, \beta\cur, \PsiMC\right)$.
\end{enumerate}

\subsection{Draw the inverse temperature parameter from $p\coh(\beta |Z)$}
Using Bayes rule, 
\begin{equation}
    p\coh(\beta |Z) \propto p\coh(Z|\beta) p\coh(\beta) ,
\label{eq:drawbeta}
\end{equation}
where $Z|\beta$ follows an Ising distribution, see
Equation~\ref{eq:ising:ising}. The partition function in Equation~\ref{eq:ising:ising}, the density of states for a periodic two-dimensional Ising lattice, is exactly calculated using Mathematica code by \cite{Beale1996Edoe}.  We sample Equation~\ref{eq:drawbeta}
via Metropolis Hastings with a gamma proposal distribuiton $J(\beta^* | \beta)$, 
$$\beta^* |\beta \sim \mbox{Gamma}\left( \frac{\beta^2}{\rho^2}, \frac{\beta}{\rho^2}\right) \; ,$$
so that $E[\beta^* |\beta ] =\beta$ and Var$[\beta^* |\beta ] = \rho$. We use a weakly informative prior on $\beta$ that favors small  values to help the Markov chain avoid the critical temperature. (If the chain reaches the critical temperature, the image `freezes,' and iterates of $\beta$ increase arbitrarily high, giving nonsensical results.) We recommend running several iterations of this Metropolis-Hastings algorithm before updating $\beta$.

\subsection{Draw the mean and variance parameters from $p(\tau_0,\tau_1,  \sigma_0^2, \sigma_1^2  | \PsiMC , Z$)}

Using Bayes rule, 
$$p\coh(\tau_0, \tau_1,  \sigma_0^2, \sigma_1^2| \PsiMC ,Z) 
\propto \prod^n_{i=1} 
p\coh(\psi_i | z_i, \tau_0, \tau_1,  \sigma_0^2, \sigma_1^2)  \,
p\coh(\tau_0,  \sigma_0^2) \,
p\coh(\tau_1, \sigma_1^2)\; ,$$ 
where $p\coh(\psi_i | z_i, \tau_0, \tau_1,  \sigma_0^2, \sigma_1^2)$ is given in Equation~\ref{eq:sqrt-lam} (recall that $\psi_i = \sqrt{\tilde\lambda_i}$)
and we specify both $p\coh(\tau_0,\sigma_0^2)$ and $p\coh(\tau_1,\sigma_1^2)$ to be $\text{Normal-Inv-}\chi^{2}$ distributions, i.e.,

\begin{eqnarray}
\sigma_1^2 &\sim& \mbox{Inverse-}\chi^2 \left( \nu_{1}, \omega_{1}^2 \right) \nonumber \\
\tau_1|\sigma_1^2 &\sim& \mbox{Normal}\left(\tau_{1}, \sigma_1^2\right) \nonumber \\
\sigma_0^2 &\sim& \mbox{Inverse-}\chi^2\left( \nu_{0}, \omega_{0}^2\right) \nonumber \\
\tau_0|\sigma_0^2 &\sim& \mbox{Normal}\left(\tau_{0}, \sigma_0^2\right) , 
\end{eqnarray}
since conditional on $Z$ this choice of prior is conjugate to $p\coh(\psi_i | z_i, \tau_0, \tau_1,  \sigma_0^2, \sigma_1^2)$. The density of the $\text{Normal-Inv-}\chi^{2}$ distribution is
\begin{align}
    \label{eq:prior on theta} 
    p\coh(\tau_k,\sigma_k^2) \propto  (\sigma_k^2)^{-(\nu+3)/2}  \exp \left( - \frac{1}{2 \sigma_k^2} [\nu \omega_k^2 + (\tau_k - \mu_k)^2]\right) ,
\end{align} 
for $k=1,2$. For simplicity we set $\mu_0=\mu_1=\mu$, $\nu_0=\nu_1=\nu$ and $\omega_0=\omega_1=\omega$.

In this case it is easily shown that the posterior distributions are  also $\text{Normal-Inv-}\chi^{2}$ distributions \citep[e.g.,][Section~3.3]{gelman1995bayesian}. Furthermore, the posterior distribution splits into two independent factors: $p(\tau_0, \tau_1,  \sigma_0^2, \sigma_1^2| \PsiMC ,Z) = p(\tau_0,  \sigma_0^2| \PsiMC ,Z)p(\tau_1,  \sigma_1^2| \PsiMC ,Z)$, which can be written in a hierarchical manner as 
\begin{eqnarray}
\sigma_1^2 |\PsiMC ,Z &\sim& \mbox{Inverse-}\chi^2 ( \nu_{n_1}, \omega_{n_1}^2) \nonumber \\
\tau_1|\sigma_1^2, \PsiMC ,Z &\sim& \mbox{Normal}(\mu_{n_1}, \sigma_1^2/(n_1+1)) \nonumber \\
\sigma_0^2 |\PsiMC ,Z &\sim& \mbox{Inverse-}\chi^2 ( \nu_{n_0}, \omega_{n_0}^2) \nonumber \\
\tau_0|\sigma_0^2, \PsiMC ,Z&\sim& \mbox{Normal}(\mu_{n_0}, \sigma_0^2/(n_0+1)) \; ,
\end{eqnarray}
where

\begin{eqnarray}
\mu_{n_1} &=& \frac{1}{1+n_1} \mu + \frac{n_1}{1+n_1} \bar \psi_{n_1}^{(t)}\nonumber \\
\nu_{n_1} &=& \nu+n_1 \nonumber \\
\nu_{n_1}\omega_{n_1} &=& \nu \omega^2 + (n_1-1)s_{n_1}^2 + \frac{n_1}{1+n_1} (\bar \psi_{n_1}^{(t)} - \mu)^2\nonumber \\
\bar \psi_{n_1}^{(t)} &=& \frac{1}{n_1}\sum_{i\in \{z_i =1\}} \psi_i^{(t)}  \nonumber \\
s_{n_1} &=&  \frac{1}{n_1-1} \sum_{i \in \{z_i=1\}} (\psi_i^{(t)} - \bar \psi_{n_1}^{(t)})^2 \nonumber \\
n_1 &=& \sum_{i=1}^n \mathbb{I}_{z_i=1} \nonumber \; ,
\end{eqnarray}
for when $z_i=+1$ and similarly defined when $z_i=-1$. As a precaution to avoid label switching, after each iteration if $\tau_1$ is less than $\tau_0$, then the $\tau_0, \tau_1$ and $\sigma_0^2, \sigma_1^2$ values are swapped for the final update.

\subsection{Draw from the posterior distribution on pixel assignments $p(Z | \PsiMC, \tau_0, \tau_1, \sigma_0^2, \sigma_1^2, \beta )$}

Introducing auxiliary variables into a MCMC scheme often enables the construction of Markov chains that mix faster and are easier to simulate. The Swendsen-Wang (SW) method is an auxiliary variable method for sampling from the Ising distribution \citep{higdon,Swendsen1987}. We define $u$ to be an auxiliary  variable with components $u_{ij}$ corresponding to each adjacency with $d(i,j)=1$. Conditional on $Z$, the $u_{ij}$ components are defined to be independent with a uniform distribution

\begin{align}
    \label{eq:uij}
    u_{ij} | Z  \sim \text{Uniform} \left(0,\exp \left(2 \beta \mathbb{I}_{z_{i} = z_{j}}\right)\right). 
\end{align} 

From Equation~\ref{eq:uij} we see that if $z_{i} = z_{j}$ then $u_{ij} | Z \sim \text{Uniform} \left(0,\exp \left(2 \beta\right)\right)$ whereas if $z_{i} \neq z_{j}$ then $u_{ij} | Z \sim \text{Uniform} (0,1)$. The conditional distribution of $Z$ given $u$ is then

\begin{align}
    \label{eq:Z|u} 
    p(Z|u) \propto \prod_{i}p\coh( \psi_{i}| z_i, \tau_0, \tau_1, \sigma_0^2, \sigma_1^2  ) \prod_{i,j: d(i,j)=1} \mathbb{I} \big{\{} 0 \leq u_{ij} \leq \exp \left(2 \beta \mathbb{I}_{z_{i} = z_{j}}\right) \big{\}} .
\end{align}
If $u_{ij}|Z > 1$, then from Equation~\ref{eq:Z|u} we see that the neighbors $z_{i}$ and $z_{j}$ must take the same value and we say that they are \textit{bonded}. The probability of a bond forming ($u_{ij} | Z > 1$) is $p = 1-e^{-2\beta}$. We can write out the probabilities for two adjacent pixel values given that they are bonded:
\begin{align}
    \label{eq:z bonded} 
    p(z_{i},z_{j} | u_{ij}>1) = \prod_{i}p\coh( \psi_{i}| z_i, \tau_0, \tau_1, \sigma_0^2, \sigma_1^2  ) \times 
    \begin{cases}
        p \quad \text{for} \quad z_{i} = z_{j} \\
        0 \quad \text{for} \quad z_{i} \neq z_{j},
    \end{cases}
\end{align}
and for the cases where they are not bonded: 
\begin{align}
    \label{eq:z not bonded} 
    p(z_{i},z_{j} | u_{ij} < 1) = \prod_{i}p\coh( \psi_{i}| z_i, \tau_0, \tau_1, \sigma_0^2, \sigma_1^2  ) \times
    \begin{cases}
        1 - p \quad \text{for} \quad z_{i} = z_{j} \\
        1 \quad \text{for} \quad z_{i} \neq z_{j}.
    \end{cases}
\end{align}

The auxiliary variable, $u$, partitions $Z$ into clusters of spins, $C \subset Z$ where every pixel in $C$ is bonded to at least one other pixel in the cluster. From Equation~\ref{eq:z bonded}, every $z_{i} \in C$ must have the same value. For each cluster, the pixel assignments are $+1$ with a probability of $p_+$ and $-1$ with a probability of $p_-=1-p_+$ where,
$$\frac{p_+}{p_-} = \frac{\prod_{i \in C} p\coh( \psi_{i} |, z_{i} = +1,\tau_1, \sigma_1^2) }{\prod_{i \in C} p\coh( \psi_{i} | z_{i} = -1, \tau_0,\sigma_0^2)} \; . $$
In the event of a pixel being in a cluster on its own we assign it to $+1$ with a probability of,
$$p_+ = \frac{p\coh( \psi_{i} |z_{i} = +1, \tau_1, \sigma_1^2) }{p\coh(\psi_{i} | z_{i} = +1, \tau_1, \sigma_1^2) +p\coh(\psi_{i}| z_{i} = -1 , \tau_0,\sigma_0^2)} \; .$$

\section{Monte Carlo Estimation of ratios of $p(Z\mid Y)$}
\label{app:Zmargin}

Integrating the posterior distribution given in Equation~\ref{eq:post-full}, we obtain
\begin{eqnarray}
p\coh(Z\mid Y) &=& \int p\coh(Z, \theta, \beta \mid  \tilde\Lambda) \; 
p\lira(\tilde\Lambda,\xi\base \mid Y) \, {\rm d}\theta \, {\rm d} \beta \, {\rm d}  \tilde\Lambda \,
{\rm d}  \tilde\xi\base\\
&=& \int p\coh(Z \mid \theta, \beta, \tilde\Lambda) \; 
p\coh(\theta, \beta \mid  \tilde\Lambda) \;
p\lira(\tilde\Lambda \mid Y) \, {\rm d}\theta \, {\rm d} \beta \, {\rm d} \tilde\Lambda.
\end{eqnarray}
Although this marginal posterior can, in principle, be evaluated via
\begin{eqnarray}
\hat p\coh(Z|Y) &= & \frac{1}{T} \sum_{t=1}^T p(Z\mid \thetaMC, \betaMC, \LamMC)  \;,
\end{eqnarray}
a numerically more stable strategy to obtain the MAP estimate involves a sequence of pairwise comparisons to determine if
$\hat p\coh(Z_1 \mid Y) > \hat p\coh(Z_2 \mid Y)$, or equivalently, 
\begin{equation}
    \frac{\hat p\coh(Z_1\mid Y)}{\hat p\coh(Z_2\mid Y)} >1 \;.
    \label{eq:ratio}
\end{equation} 
The ratio in Equation~\ref{eq:ratio} can be computed stably via

$$\begin{aligned}
&\frac{\hat p\coh(Z_1 \mid Y)}{\hat p\coh(Z_2\mid Y)} 
=\frac{\sum_{t=1}^T \exp(\log  \;p\coh(Z_1 \mid \thetaMC, \betaMC, \LamMC))}{\sum_{t=1}^T \exp(\log \; p\coh(Z_2 \mid \thetaMC, \betaMC, \LamMC))} \nonumber \\
&\\
&= \frac{\sum_{t=1}^T \exp\Big(\log \; p\coh(Z_1 \mid \thetaMC, \betaMC,  \LamMC) -\log \; p\coh(Z_2 \mid \thetaMC, \betaMC, \LamMC)\Big)  \ \exp\Big(\log \; p\coh(Z_2 \mid \thetaMC, \betaMC, \LamMC) - l_{\mbox{max}}\Big)}
{\sum_{t=1}^T \exp\Big(\log \; p\coh(Z_2 \mid \thetaMC, \betaMC, \LamMC) - l_{\mbox{max}}\Big)} \nonumber  \\
&\\
&= \sum_{t=1}^T w_t \exp\left(\log\left(\frac{p\coh(Z_1 \mid  \thetaMC, \betaMC, \LamMC)}{p\coh(Z_2 \mid \thetaMC, \betaMC, \LamMC)}\right) \right)  \; , \nonumber
\end{aligned}$$
where 

\begin{align}
    \label{eq:w_t} 
w_t =\frac{\exp(\log \; p\coh(Z_2 \mid \thetaMC, \betaMC, \LamMC)- l_{\mbox{max}})}{  \sum_{t=1}^T \exp(\log \;p\coh(Z_2 \mid \thetaMC, \betaMC,  \LamMC) - l_{\mbox{max}}) } \nonumber
\end{align} 
and $l_{\mbox{max}} = \max_t \log \; p\coh(Z_2\mid \thetaMC, \betaMC,\LamMC)$ is the maximum log-likelihood of the denominator term.

\FloatBarrier
\newpage
\bibliography{main}

@ARTICLE{Fabbiano2019,
       author = {{Fabbiano}, G. and {Siemiginowska}, A. and {Paggi}, A. and {Elvis}, M. and {Volonteri}, M. and {Mayer}, L. and {Karovska}, M. and {Maksym}, W.~P. and {Risaliti}, G. and {Wang}, Junfeng},
        title = "{Deep Chandra Observations of ESO 428-G014. IV. The Morphology of the Nuclear Region in the Hard Continuum and Fe K{\ensuremath{\alpha}} Line}",
      journal = {\apj},
         year = 2019,
        month = jan,
       volume = {870},
       number = {2},
          eid = {69},
        pages = {69},
          doi = {10.3847/1538-4357/aaf0a4},
archivePrefix = {arXiv},
       eprint = {1811.06436},
 primaryClass = {astro-ph.GA},
       adsurl = {https://ui.adsabs.harvard.edu/abs/2019ApJ...870...69F}
}

@ARTICLE{Reddy2023,
       author = {{Reddy}, Karthik and {Georganopoulos}, Markos and {Meyer}, Eileen T. and {Keenan}, Mary and {Kollmann}, Kassidy E.},
        title = "{Offsets between X-Ray and Radio Components in X-Ray Jets: The AtlasX}",
      journal = {\apjs},
         year = 2023,
        month = mar,
       volume = {265},
       number = {1},
          eid = {8},
        pages = {8},
          doi = {10.3847/1538-4365/aca321},
archivePrefix = {arXiv},
       eprint = {2212.02061},
 primaryClass = {astro-ph.HE},
       adsurl = {https://ui.adsabs.harvard.edu/abs/2023ApJS..265....8R}
}

@ARTICLE{Reddy2021,
       author = {{Reddy}, Karthik and {Georganopoulos}, Markos and {Meyer}, Eileen T.},
        title = "{X-Ray-to-radio Offset Inference from Low-count X-Ray Jets}",
      journal = {\apjs},
         year = 2021,
        month = apr,
       volume = {253},
       number = {2},
          eid = {37},
        pages = {37},
          doi = {10.3847/1538-4365/abd8d7},
archivePrefix = {arXiv},
       eprint = {2101.02024},
 primaryClass = {astro-ph.HE},
       adsurl = {https://ui.adsabs.harvard.edu/abs/2021ApJS..253...37R}
}

@INPROCEEDINGS{Donath2022,
       author = {{Donath}, Axel and {Siemiginowska}, Aneta and {Kashyap}, Vinay and {Burke}, Douglas and {Reddy}, Karthik and {Van Dyk}, David},
        title = "{Pylira: deconvolution of images in the presence of Poisson noise}",
    booktitle = {Proceedings of the 21st Python in Science Conference},
         year = 2022,
        month = jul,
        pages = {98-104},
          doi = {10.25080/majora-212e5952-00f},
       adsurl = {https://ui.adsabs.harvard.edu/abs/2022pysc.conf...98D}
}

@book{ChandraBook,
editor = {Wilkes, Belinda and Tucker, Wallace},
title = {The Chandra X-ray Observatory},
publisher = {IOP Publishing},
year = {2019},
series = {2514-3433},
isbn = {978-0-7503-2163-1},
url = {https://dx.doi.org/10.1088/2514-3433/ab43dc},
doi = {10.1088/2514-3433/ab43dc}
}

@ARTICLE{Snios2022,
       author = {{Snios}, Bradford and {Schwartz}, Daniel A. and {Siemiginowska}, Aneta and {Sobolewska}, Ma{\l}gosia and {Birkinshaw}, Mark and {Cheung}, C.~C. and {Gobeille}, Doug B. and {Marshall}, Herman L. and {Migliori}, Giulia and {Wardle}, John F.~C. and {Worrall}, Diana M.},
        title = "{X-Ray Jets in the High-redshift Quasars J1405+0415 and J1610+1811}",
      journal = {\apj},
         year = 2022,
        month = aug,
       volume = {934},
       number = {2},
          eid = {107},
        pages = {107},
          doi = {10.3847/1538-4357/ac7cf2},
       adsurl = {https://ui.adsabs.harvard.edu/abs/2022ApJ...934..107S}
}

@INPROCEEDINGS{Fruscione2006,
       author = {{Fruscione}, Antonella and {McDowell}, Jonathan C. and {Allen}, Glenn E. and {Brickhouse}, Nancy S. and {Burke}, Douglas J. and {Davis}, John E. and {Durham}, Nick and {Elvis}, Martin and {Galle}, Elizabeth C. and {Harris}, Daniel E. and {Huenemoerder}, David P. and {Houck}, John C. and {Ishibashi}, Bish and {Karovska}, Margarita and {Nicastro}, Fabrizio and {Noble}, Michael S. and {Nowak}, Michael A. and {Primini}, Frank A. and {Siemiginowska}, Aneta and {Smith}, Randall K. and {Wise}, Michael},
        title = "{CIAO: Chandra's data analysis system}",
    booktitle = {Observatory Operations: Strategies, Processes, and Systems},
         year = 2006,
       editor = {{Silva}, David R. and {Doxsey}, Rodger E.},
       series = {Society of Photo-Optical Instrumentation Engineers (SPIE) Conference Series},
       volume = {6270},
        month = jun,
          eid = {62701V},
        pages = {62701V},
          doi = {10.1117/12.671760},
       adsurl = {https://ui.adsabs.harvard.edu/abs/2006SPIE.6270E..1VF}
}

@ARTICLE{Siemignowska2024,
       author = {{Siemiginowska}, Aneta and {Burke}, Douglas and {G{\"u}nther}, Hans Moritz and {Lee}, Nicholas P. and {McLaughlin}, Warren and {Principe}, David A. and {Cheer}, Harlan and {Fruscione}, Antonella and {Laurino}, Omar and {McDowell}, Jonathan and {Terrell}, Marie},
        title = "{Sherpa: An Open-source Python Fitting Package}",
      journal = {\apjs},
         year = 2024,
        month = oct,
       volume = {274},
       number = {2},
          eid = {43},
        pages = {43},
          doi = {10.3847/1538-4365/ad7bab},
archivePrefix = {arXiv},
       eprint = {2409.10400},
 primaryClass = {astro-ph.IM},
       adsurl = {https://ui.adsabs.harvard.edu/abs/2024ApJS..274...43S}
}

@INPROCEEDINGS{Freeman2001,
       author = {{Freeman}, Peter and {Doe}, Stephen and {Siemiginowska}, Aneta},
        title = "{Sherpa: a mission-independent data analysis application}",
    booktitle = {Astronomical Data Analysis},
         year = 2001,
       editor = {{Starck}, Jean-Luc and {Murtagh}, Fionn D.},
       series = {Society of Photo-Optical Instrumentation Engineers (SPIE) Conference Series},
       volume = {4477},
        month = nov,
        pages = {76-87},
          doi = {10.1117/12.447161},
archivePrefix = {arXiv},
       eprint = {astro-ph/0108426},
 primaryClass = {astro-ph},
       adsurl = {https://ui.adsabs.harvard.edu/abs/2001SPIE.4477...76F}
}

@ARTICLE{DeLaney2010,
       author = {{DeLaney}, Tracey and {Rudnick}, Lawrence and {Stage}, M.~D. and {Smith}, J.~D. and {Isensee}, Karl and {Rho}, Jeonghee and {Allen}, Glenn E. and {Gomez}, Haley and {Kozasa}, Takashi and {Reach}, William T. and {Davis}, J.~E. and {Houck}, J.~C.},
        title = "{The Three-dimensional Structure of Cassiopeia A}",
      journal = {\apj},
         year = 2010,
        month = dec,
       volume = {725},
       number = {2},
        pages = {2038-2058},
          doi = {10.1088/0004-637X/725/2/2038},
archivePrefix = {arXiv},
       eprint = {1011.3858},
 primaryClass = {astro-ph.GA},
       adsurl = {https://ui.adsabs.harvard.edu/abs/2010ApJ...725.2038D}
}

@ARTICLE{Vink2012,
       author = {{Vink}, Jacco},
        title = "{Supernova remnants: the X-ray perspective}",
      journal = {\aapr},
         year = 2012,
        month = dec,
       volume = {20},
          eid = {49},
        pages = {49},
          doi = {10.1007/s00159-011-0049-1},
archivePrefix = {arXiv},
       eprint = {1112.0576},
 primaryClass = {astro-ph.HE},
       adsurl = {https://ui.adsabs.harvard.edu/abs/2012A&ARv..20...49V}
}

@ARTICLE{Fabian2006,
       author = {{Fabian}, A.~C. and {Sanders}, J.~S. and {Taylor}, G.~B. and {Allen}, S.~W. and {Crawford}, C.~S. and {Johnstone}, R.~M. and {Iwasawa}, K.},
        title = "{A very deep Chandra observation of the Perseus cluster: shocks, ripples and conduction}",
      journal = {\mnras},
         year = 2006,
        month = feb,
       volume = {366},
       number = {2},
        pages = {417-428},
          doi = {10.1111/j.1365-2966.2005.09896.x},
archivePrefix = {arXiv},
       eprint = {astro-ph/0510476},
 primaryClass = {astro-ph},
       adsurl = {https://ui.adsabs.harvard.edu/abs/2006MNRAS.366..417F}
}

@ARTICLE{Fabian2012,
       author = {{Fabian}, A.~C.},
        title = "{Observational Evidence of Active Galactic Nuclei Feedback}",
      journal = {\araa},
         year = 2012,
        month = sep,
       volume = {50},
        pages = {455-489},
          doi = {10.1146/annurev-astro-081811-125521},
archivePrefix = {arXiv},
       eprint = {1204.4114},
 primaryClass = {astro-ph.CO},
       adsurl = {https://ui.adsabs.harvard.edu/abs/2012ARA&A..50..455F}
}

@article{worr:etal:20,
    author = {Worrall, D M and Birkinshaw, M and Marshall, H L and Schwartz, D A and Siemiginowska, A and Wardle, J F C},
    title = {Inverse-Compton scattering in the resolved jet of the high-redshift quasar PKS J1421â0643},
    journal = {Monthly Notices of the Royal Astronomical Society},
    volume = {497},
    number = {1},
    pages = {988-1000},
    year = {2020},
    month = {07}
}

@article{meye:etal21,
  	title={e{BASCS}: Disentangling overlapping astronomical sources {II}, using spatial, spectral, and temporal information},
  	author={Meyer, Antoine D and van~Dyk, David A and Kashyap, Vinay L and Campos, Luis F and 
		Jones, David E and Siemiginowska, Aneta and Zezas, Andreas},
  	journal={Monthly Notices of the Royal Astronomical Society},
  	volume={506},
  	number={4},
  	pages={6160--6180},
  	year={2021},
  	publisher={Oxford University Press}
}

@ARTICLE{nowa:kola:00,
  author={Nowak, R.D. and Kolaczyk, E.D.},
  journal={IEEE Transactions on Information Theory}, 
  title={A statistical multiscale framework for Poisson inverse problems}, 
  year={2000},
  volume={46},
  number={5},
  pages={1811-1825},
  doi={10.1109/18.857793}}

@article{sten:etal:18ii,
  	title={{B}ayesian Statistical Methods For Astronomy {Part II: M}arkov Chain {M}onte {C}arlo},
  	author={Stenning, David C and van~Dyk, David A},
  	journal={Statistics for Astrophysics: Bayesian Methodology},
  	pages={29-58},
  	year={2018},
  	publisher={EDP Sciences}
}

@article{Esch2004,
author = {Esch, David N. and Connors, Alanna and Karovska, Margarita and van Dyk, David A.},
doi = {10.1086/421761},
issn = {0004-637X},
journal = {The Astrophysical Journal},
number = {2},
pages = {1213--1227},
title = {{An Image Restoration Technique with Error Estimates}},
url = {http://stacks.iop.org/0004-637X/610/i=2/a=1213},
volume = {610},
year = {2004}
}

@article{Connors2007,
author = {Connors, A. and van Dyk, D. A.},
journal = {Statistical Challenges in Modern Astronomy IV},
pages = {101--117},
title = {{How to win with non-Gaussian data: Poisson goodness-of-fit}},
url = {http://aspbooks.org/custom/publications/paper/371-0101.html},
volume = {371},
year = {2007}
}

@article{McKeough2016,
	doi = {10.3847/1538-4357/833/1/123},
	url = {https://doi.org/10.3847%2F1538-4357%2F833%2F1%2F123},
	year = 2016,
	month = {dec},
	publisher = {American Astronomical Society},
	volume = {833},
	number = {1},
	pages = {123},
	author = {Kathryn McKeough and Aneta Siemiginowska and C. C. Cheung and {\L}ukasz Stawarz and Vinay L. Kashyap and Nathan Stein and Vasileios Stampoulis and David A. van Dyk and J. F. C. Wardle and N. P. Lee and D. E. Harris and D. A. Schwartz and Davide Donato and Laura Maraschi and Fabrizio Tavecchio},
	title = {Detecting Relativistic X-Ray Jets in High -Redshift Quasars},
	journal = {The Astrophysical Journal}
}

@article{Stein2015,
archivePrefix = {arXiv},
arxivId = {1510.04662},
author = {Stein, Nathan M. and {Van Dyk}, David A. and Kashyap, Vinay L. and Siemiginowska, Aneta},
doi = {10.1088/0004-637X/813/1/66},
eprint = {1510.04662},
issn = {15384357},
journal = {Astrophysical Journal},
number = {1},
pages = {66},
publisher = {IOP Publishing},
title = {{Detecting unspecified structure in low-count images}},
url = {http://dx.doi.org/10.1088/0004-637X/813/1/66},
volume = {813},
year = {2015}
}

@article{Swendsen1987,
issn = {00319007},
journal = {Physical review letters},
pages = {86},
volume = {58},
number = {2},
year = {1987},
title = {Nonuniversal critical dynamics in Monte Carlo simulations},
language = {eng},
author = {Swendsen and Wang},
}

@article{ising1925,
journal = {Physik},
pages = {253-258},
volume = {31},
year = {1925},
title = {Beitrag zur Theorie des Ferromagnetismus},
language = {ger},
author = {Ising, E.},
}

@article{potts_1952, 
title={Some generalized order-disorder transformations},
volume={48}, 
DOI={10.1017/S0305004100027419}, 
number={1}, 
journal={Mathematical Proceedings of the Cambridge Philosophical Society}, publisher={Cambridge University Press}, 
author={Potts, R. B.}, 
year={1952}, 
pages={106–109}}

@article{Beale1996Edoe,
issn = {00319007},
journal = {Physical review letters},
pages = {78},
volume = {76},
number = {1},
year = {1996},
title = {Exact distribution of energies in the two-dimensional ising model},
language = {eng},
author = {Beale},
}

@article{Kashyap2017,
author = {Kashyap, Vinay L. and {Van Dyk}, David and McKeough, Katy and Primini, Frank and Jerius, Diab and Gowrishankar, Akshay and Siemiginowska, Aneta},
doi = {10.1017/S1743921317004690},
issn = {17439221},
journal = {Proceedings of the International Astronomical Union},
number = {S331},
pages = {284--289},
title = {{X-raying the evolution of SN 1987A}},
volume = {12},
year = {2017}
}

@article{Marquez-Neila2014,
author = {Marquez-Neila, Pablo and Baumela, Luis and Alvarez, Luis},
doi = {10.1109/TPAMI.2013.106},
isbn = {0162-8828},
issn = {01628828},
journal = {IEEE Transactions on Pattern Analysis and Machine Intelligence},
number = {1},
pages = {2--17},
pmid = {24231862},
title = {{A morphological approach to curvature-based evolution of curves and surfaces}},
volume = {36},
year = {2014}
}

@article{Adams1994,
author = {Adams, Rolf and Bischof, Leanne},
issn = {0162-8828},
journal = {IEEE Transactions on Pattern Analysis and Machine Intelligence},
number = {6},
pages = {641--647},
title = {{Seeded Region growing}},
volume = {16},
year = {1994}
}

@article{Bentrem2010,
author = {Bentrem, Frank W.},
doi = {10.2478/s11534-009-0165-y},
isbn = {1153400901},
issn = {18951082},
journal = {Central European Journal of Physics},
number = {5},
pages = {689--698},
title = {{A Q-Ising model application for linear-time image segmentation}},
volume = {8},
year = {2010}
}

@article{Mignotte2000a,
author = {Mignotte, Max and Collet, Christophe and P{\'{e}}rez, Patrick and Bouthemy, Patrick},
doi = {10.1109/83.847834},
issn = {10577149},
journal = {IEEE Transactions on Image Processing},
number = {7},
pages = {1216--1231},
title = {{Sonar image segmentation using an unsupervised hierarchical MRF model}},
volume = {9},
year = {2000}
}

@article{Sanders2001,
archivePrefix = {arXiv},
arxivId = {astro-ph/0011500},
author = {Sanders, J. S. and Fabian, A. C.},
doi = {10.1046/j.1365-8711.2001.04410.x},
eprint = {0011500},
issn = {00358711},
journal = {Monthly Notices of the Royal Astronomical Society},
number = {1},
pages = {178--186},
primaryClass = {astro-ph},
title = {{Adaptive binning of X-ray galaxy cluster images}},
volume = {325},
year = {2001}
}

@article{Sanders2006,
archivePrefix = {arXiv},
arxivId = {astro-ph/0606528},
author = {Sanders, J S},
doi = {10.1111/j.1365-2966.2006.10716.x},
eprint = {0606528},
issn = {00358711},
journal = {Monthly Notices of the Royal Astronomical Society},
number = {2},
pages = {829--842},
primaryClass = {astro-ph},
title = {{Contour binning: A new technique for spatially resolved X-ray spectroscopy applied to Cassiopeia A}},
volume = {371},
year = {2006}
}

@article{Vikhlinin1998,
author = {Vikhlinin, A. and McNamara, B. R. and Forman, W. and Jones, C. and Quintana, H. and Hornstrup, A.},
doi = {10.1086/305951},
issn = {0004-637X},
journal = {The Astrophysical Journal},
number = {2},
pages = {558--581},
title = {{A Catalog of 203 Galaxy Clusters Serendipitously Detected in the ROSAT PSPC Pointed Observations}},
volume = {502},
year = {1998}
}

@article{fan:etal:23,
  	title={Identifying diffuse spatial structures in high-energy photon lists},
  	author={Fan, Minjie and Wang, Jue and Kashyap, Vinay L and Lee, Thomas CM and van~Dyk, David A and Zezas, Andreas},
  	journal={The Astronomical Journal},
  	volume={165},
  	number={2},
  	pages={66},
  	year={2023},
  	publisher={IOP Publishing}
}

@ARTICLE{wang:etal:24,
       author = {{Wang}, Jue and {Kashyap}, Vinay L. and {Lee}, Thomas C.~M. and {van Dyk}, David A. and {Zezas}, Andreas},
        title = "{Auto-BUQ: Uncertainty Quantification for the Boundaries of Segmented Events}",
      journal = {\aj},
         year = 2025,
        month = jun,
       volume = {169},
       number = {6},
          eid = {329},
        pages = {329},
          doi = {10.3847/1538-3881/adc931},
       adsurl = {https://ui.adsabs.harvard.edu/abs/2025AJ....169..329W}
}

@article{Jones2014,
doi = {10.1088/0004-637x/808/2/137},
	url = {https://doi.org/10.1088%2F0004-637x%2F808%2F2%2F137},
	year = 2015,
	publisher = {{IOP} Publishing},
	volume = {808},
	number = {2},
	pages = {137},
	author = {David E. Jones and Vinay L. Kashyap and David A. van Dyk},
	title = {Disentangling Overlapping Astronomical Sources Using Spatial and Spectral Information},
	journal = {The Astrophysical Journal}
}

@article{Picquenot2019,
archivePrefix = {arXiv},
arxivId = {1905.10175},
author = {Picquenot, A. and Acero, F. and Bobin, J. and Maggi, P. and Ballet, J. and Pratt, G. W.},
doi = {10.1051/0004-6361/201834933},
eprint = {1905.10175},
issn = {14320746},
journal = {Astronomy and Astrophysics},
title = {{Novel method for component separation of extended sources in X-ray astronomy}},
volume = {627},
year = {2019}
}

@article{Starck2002,
author = {Starck, J L and Pantin, E and Murtagh, Fionn},
journal = {Publications of the Astronomical Society of the Pacific},
number = {800},
pages = {1051--1069},
title = {{Deconvolution in astronomy: a review}},
url = {http://www.journals.uchicago.edu/PASP/home.html},
volume = {114},
year = {2002}
}

@article{Freeman2002,
author = {Freeman, Peter E. and Kashyap, Vinay and Rosner, R and Lamb, D.Q.},
journal = {The Astrophysical Journal Supplement Series},
number = {138},
pages = {185--218},
title = {{A wavelet-based algorithm for the spatial analysis of poisson data}},
year = {2002}
}

@article{Ebeling1993,
author = {Ebeling, H. and Wiedenmann, G.},
doi = {10.1103/PhysRevE.47.704},
issn = {1063651X},
journal = {Physical Review E},
number = {1},
pages = {704--710},
title = {{Detecting structure in two dimensions combining Voronoi tessellation and percolation}},
volume = {47},
year = {1993}
}

@article{Ebeling2006,
archivePrefix = {arXiv},
arxivId = {astro-ph/0601306},
author = {Ebeling, H. and White, D. A. and Rangarajan, F. V. N.},
doi = {10.1111/j.1365-2966.2006.10135.x},
eprint = {0601306},
issn = {0035-8711},
journal = {Monthly Notices of the Royal Astronomical Society},
number = {1},
pages = {65--73},
primaryClass = {astro-ph},
title = {{ASMOOTH: a simple and efficient algorithm for adaptive kernel smoothing of two-dimensional imaging data}},
volume = {368},
year = {2006}
}

@article{Bertin1996,
author = {Bertin, E. and Arnouts, S.},
doi = {10.1051/aas:1996164},
issn = {03650138},
journal = {Astronomy and Astrophysics Supplement Series},
number = {2},
pages = {393--404},
title = {{SExtractor: Software for source extraction}},
volume = {117},
year = {1996}
}

@article{Gonzalez-Gaitan2019,
archivePrefix = {arXiv},
arxivId = {1802.06280},
author = {Gonz{\'{a}}lez-Gait{\'{a}}n, S. and {De Souza}, R. S. and Krone-Martins, A. and Cameron, E and Coelho, P and Galbany, L and Ishida, E. E.O.},
doi = {10.1093/mnras/sty2881},
eprint = {1802.06280},
issn = {13652966},
journal = {Monthly Notices of the Royal Astronomical Society},
number = {3},
pages = {3880--3891},
title = {{Spatial field reconstruction with INLA: Application to IFU galaxy data}},
volume = {482},
year = {2019}
}

@article{Beasley_2002,
	doi = {10.1086/339806},
	url = {https://doi.org/10.1086%2F339806},
	year = 2002,
	month = {jul},
	publisher = {{IOP} Publishing},
	volume = {141},
	number = {1},
	pages = {13--21},
	author = {A. J. Beasley and D. Gordon and A. B. Peck and L. Petrov and D. S. MacMillan and E. B. Fomalont and C. Ma},
	title = {The {VLBA} Calibrator Survey{\textemdash}{VCS}1},
	journal = {The Astrophysical Journal Supplement Series}
}

@article{ellison2004,
author = {Ellison, S.L and L., Yan and Hook, I.M. and Pettini, M. and Wall, J.V. and Shaver, P.},
doi = {10.1051/0004-6361},
issn = {0004-6361},
journal = {Astronomy and Astrophysics},
pages = {393--406},
title = {{The CORALS survey I: New estimates of the number density and gas content of damped Lyman alpha systems free from dust bias}},
volume = {379},
year = {2001}
}

@article{higdon,
  title={Auxiliary variable methods for Markov chain Monte Carlo with applications},
  author={Higdon, David M},
  journal={Journal of the American statistical Association},
  volume={93},
  number={442},
  pages={585--595},
  year={1998},
  publisher={Taylor \& Francis}
}

@book{gelman1995bayesian,
  title={Bayesian data analysis},
  author={Gelman, Andrew and Carlin, John B and Stern, Hal S and Rubin, Donald B},
  year={1995},
  publisher={Chapman and Hall/CRC}
}

@ARTICLE{Harris2006,
       author = {{Harris}, D.~E. and {Krawczynski}, Henric},
        title = "{X-Ray Emission from Extragalactic Jets}",
      journal = {\araa},
         year = 2006,
        month = sep,
       volume = {44},
       number = {1},
        pages = {463-506},
          doi = {10.1146/annurev.astro.44.051905.092446},
archivePrefix = {arXiv},
       eprint = {astro-ph/0607228},
 primaryClass = {astro-ph},
       adsurl = {https://ui.adsabs.harvard.edu/abs/2006ARA&A..44..463H}
}

@INPROCEEDINGS{Cheung2008,
       author = {{Cheung}, C.~C. and {Stawarz}, L. and {Siemiginowska}, A. and {Harris}, D.~E. and {Schwartz}, D.~A. and {Wardle}, J.~F.~C. and {Gobeille}, D. and {Lee}, N.~P.},
        title = "{The Highest Redshift Relativistic Jets}",
    booktitle = {Extragalactic Jets: Theory and Observation from Radio to Gamma Ray},
         year = 2008,
       editor = {{Rector}, T.~A. and {De Young}, D.~S.},
       series = {Astronomical Society of the Pacific Conference Series},
       volume = {386},
        month = jun,
        pages = {462},
          doi = {10.48550/arXiv.0712.1192},
archivePrefix = {arXiv},
       eprint = {0712.1192},
 primaryClass = {astro-ph},
       adsurl = {https://ui.adsabs.harvard.edu/abs/2008ASPC..386..462C}
}

@ARTICLE{Plsek2024,
       author = {{Pl{\v{s}}ek}, T. and {Werner}, N. and {Topinka}, M. and {Simionescu}, A.},
        title = "{CAvity DEtection Tool (CADET): pipeline for detection of X-ray cavities in hot galactic and cluster atmospheres}",
      journal = {\mnras},
         year = 2024,
        month = jan,
       volume = {527},
       number = {2},
        pages = {3315-3346},
          doi = {10.1093/mnras/stad3371},
archivePrefix = {arXiv},
       eprint = {2304.05457},
 primaryClass = {astro-ph.HE},
       adsurl = {https://ui.adsabs.harvard.edu/abs/2024MNRAS.527.3315P}
}

@ARTICLE{Borlaff2024,
       author = {{Borlaff}, Alejandro S. and {Marcum}, Pamela M. and {Alpaslan}, Mehmet and {Temi}, Pasquale and {Chamba}, Nushkia and {Chojnowski}, Drew S. and {Fanelli}, Michael N. and {Koekemoer}, Anton M. and {Laine}, Seppo and {Lopez-Rodriguez}, Enrique and {Siemiginowska}, Aneta},
        title = "{SAUNAS. I. Searching for Low Surface Brightness X-Ray Emission with Chandra/ACIS}",
      journal = {\apj},
         year = 2024,
        month = jun,
       volume = {967},
       number = {2},
          eid = {169},
        pages = {169},
          doi = {10.3847/1538-4357/ad3c37},
archivePrefix = {arXiv},
       eprint = {2405.01625},
 primaryClass = {astro-ph.GA},
       adsurl = {https://ui.adsabs.harvard.edu/abs/2024ApJ...967..169B}
}

@ARTICLE{Borlaff2024b,
       author = {{Borlaff}, Alejandro S. and {Marcum}, Pamela M. and {Temi}, Pasquale and {Chamba}, Nushkia and {Chojnowski}, S. Drew and {Lopez-Rodriguez}, Enrique and {Siemiginowska}, Aneta and {Laine}, Seppo and {Koekemoer}, Anton M. and {Sanderson}, Kelly N. and {Dijeau}, Audrey F. and {Prescott}, Moire K.~M. and {Proudfit}, Leslie and {Fanelli}, Michael N.},
        title = "{SAUNAS. II. Discovery of Cross-shaped X-Ray Emission and a Rotating Circumnuclear Disk in the Supermassive S0 Galaxy NGC 5084}",
      journal = {\apj},
         year = 2024,
        month = dec,
       volume = {977},
       number = {2},
          eid = {238},
        pages = {238},
          doi = {10.3847/1538-4357/ad7c4b},
archivePrefix = {arXiv},
       eprint = {2408.10449},
 primaryClass = {astro-ph.GA},
       adsurl = {https://ui.adsabs.harvard.edu/abs/2024ApJ...977..238B}
}

@ARTICLE{Chamba2025,
       author = {{Chamba}, Nushkia and {Marcum}, Pamela M. and {Borlaff}, Alejandro S. and {Temi}, Pasquale and {Siemiginowska}, Aneta},
        title = "{Truncations in the X-Ray Halos of Early-type Galaxies as a Tracer of Feedback and Mergers}",
      journal = {\apj},
         year = 2025,
        month = aug,
       volume = {988},
       number = {2},
          eid = {249},
        pages = {249},
          doi = {10.3847/1538-4357/ade873},
archivePrefix = {arXiv},
       eprint = {2506.14884},
 primaryClass = {astro-ph.GA},
       adsurl = {https://ui.adsabs.harvard.edu/abs/2025ApJ...988..249C}
}

@ARTICLE{Siemiginowska2007,
       author = {{Siemiginowska}, Aneta and {Stawarz}, {\L}ukasz and {Cheung}, C.~C. and {Harris}, D.~E. and {Sikora}, Marek and {Aldcroft}, Thomas L. and {Bechtold}, Jill},
        title = "{The 300 kpc Long X-Ray Jet in PKS 1127-145, z = 1.18 Quasar: Constraining X-Ray Emission Models}",
      journal = {\apj},
         year = 2007,
        month = mar,
       volume = {657},
       number = {1},
        pages = {145-158},
          doi = {10.1086/510898},
archivePrefix = {arXiv},
       eprint = {astro-ph/0611406},
 primaryClass = {astro-ph},
       adsurl = {https://ui.adsabs.harvard.edu/abs/2007ApJ...657..145S}
}

@ARTICLE{Hardcastle2016,
       author = {{Hardcastle}, M.~J. and {Lenc}, E. and {Birkinshaw}, M. and {Croston}, J.~H. and {Goodger}, J.~L. and {Marshall}, H.~L. and {Perlman}, E.~S. and {Siemiginowska}, A. and {Stawarz}, {\L}. and {Worrall}, D.~M.},
        title = "{Deep Chandra observations of Pictor A}",
      journal = {\mnras},
         year = 2016,
        month = feb,
       volume = {455},
       number = {4},
        pages = {3526-3545},
          doi = {10.1093/mnras/stv2553},
archivePrefix = {arXiv},
       eprint = {1510.08392},
 primaryClass = {astro-ph.HE},
       adsurl = {https://ui.adsabs.harvard.edu/abs/2016MNRAS.455.3526H}
}
\bibliographystyle{aasjournalv7}



\end{document}